\documentclass[sigconf]{acmart}
\AtBeginDocument{%
  }

\usepackage{multirow}
\usepackage{makecell}
\usepackage[table]{xcolor}
\usepackage{bbm}
\usepackage{algorithm}
\usepackage{algorithmic}

\setcopyright{acmlicensed}
\copyrightyear{2026}
\acmYear{2026}
\acmDOI{XXXXXXX.XXXXXXX}
\acmConference[MM'26]{the 34th ACM International Conference on Multimedia}{November 10–-14, 2026}{Rio de Janeiro, Brazil}
\acmISBN{978-1-4503-XXXX-X/2018/06}

\copyrightyear{2026}
\acmYear{2026}
\setcopyright{cc}
\setcctype{by}
\acmConference[MM '26]{Proceedings of the 34th ACM International Conference on Multimedia}{November 10--14, 2026}{Rio de Janeiro, Brazil}
\acmBooktitle{Proceedings of the 34th ACM International Conference on Multimedia (MM '26), November 10--14, 2026, Rio de Janeiro, Brazil}
\acmDOI{10.1145/3767308.3836330}
\acmISBN{979-8-4007-2213-4/2026/11}

\begin{document}


\title{M$^3$Prune: Hierarchical Collaborative Pruning for Efficient Multi-Modal Multi-Agent Retrieval-Augmented Generation}

\author{Taolin Zhang}
\authornote{Both authors contributed equally to this research.}
\email{tlzhang@hfut.edu.cn}
\affiliation{%
  \institution{Hefei University of Technology}
  \city{Hefei}
  \country{China}
}

\author{Weizi Shao}
\authornotemark[1]
\affiliation{%
  \institution{East China Normal University}
  \city{Shanghai}
  \country{China}}
\email{51265901007@stu.ecnu.edu.cn}


\author{Zijie Zhou}
\affiliation{%
  \institution{China University of Petroleum}
  \city{Beijing}
  \country{China}}
\email{zjzhouzh@gmail.com}

\author{Chen Chen}
\affiliation{%
  \institution{Guangdong University of Finance and Economics}
  \city{Guangdong}
  \country{China}}
\email{allen821@student.gdufe.edu.cn}

\author{Daiyang Yu}
\affiliation{%
  \institution{Hefei University of Technology}
  \city{Hefei}
  \country{China}}
\email{2023216618@mail.hfut.edu.cn}

\author{Tingyuan Hu}
\affiliation{%
  \institution{East China Normal University}
  \city{Shanghai}
  \country{China}}
\email{10245102409@stu.ecnu.edu.cn}

\author{Chengyu Wang}
\authornote{Corresponding Author.}
\affiliation{%
  \institution{Alibaba Group}
  \city{Hangzhou}
  \country{China}}
\email{chengyu.wcy@alibaba-inc.com}


\author{Xiaofeng He}
\affiliation{%
  \institution{East China Normal University}
  \city{Shanghai}
  \country{China}}
\email{hexf@cs.ecnu.edu.cn}

\begin{abstract}
Recent advances in multi-modal retrieval-augmented generation (mRAG), which augments multi-modal large language models (MLLMs) with external knowledge, have shown that collective intelligence from multiple agents can outperform a single model through effective communication.
Despite their strong performance, existing multi-agent systems incur substantial token overhead and computational cost, posing challenges for large-scale deployment.
To address these issues, we propose a \textbf{M}ulti-\textbf{M}odal \textbf{M}ulti-agent hierarchical communication graph \textbf{PRUNING} framework, termed \textbf{M$^3$Prune}.
M$^3$Prune eliminates redundant communication edges both across and within modalities, improving the trade-off between task performance and token overhead.
Specifically, M$^3$Prune first performs intra-modal graph sparsification in the textual and visual modalities to identify task-critical communication links.
It then constructs an inter-modal communication graph and sparsifies cross-modal connections while encouraging consistent cross-modal reasoning through a modality alignment score.
Finally, it progressively prunes redundant edges to obtain an efficient hierarchical topology.
Extensive experiments on both general-domain and domain-specific mRAG benchmarks show that M$^3$Prune consistently outperforms single-agent and strong multi-agent mRAG systems while significantly improving token efficiency. The code and data are available at \url{https://github.com/ztl-35/M3Prune}.
\end{abstract}

\ccsdesc[500]{methodologies~Artificial intelligence}

\keywords{Multi-Modal Retrieval-Augmented Generation; Multi-Modal Multi-Agent Systems; Graph Optimization}



\maketitle

\section{Introduction}
While RAG has achieved significant success in the textual domain~\cite{DBLP:conf/acl/0001LC0H25,DBLP:conf/acl/Lee0MHAIRF25,DBLP:conf/acl/0002JRPYWHX0D025}, extending it to the multi-modal setting remains challenging~\cite{DBLP:conf/naacl/JiangXWYHBSTM25}.
Traditional methods typically rely on a cascaded pipeline in which a multi-modal retriever extracts relevant evidence, which is then forwarded to LLMs for answer synthesis~\cite{DBLP:conf/cvpr/KhanKDGLS21,DBLP:conf/iccv/QianWDQL023,DBLP:conf/iccv/HuLKWOKS23}.
This segmented approach inherently suffers from global semantic misalignment~\cite{DBLP:conf/wacv/LinBSSB23,DBLP:journals/corr/abs-2509-25638}.

The emergence of multi-modal large language models (MLLMs) presents a transformative opportunity, as these models exhibit strong capabilities for modeling vision--language correlations~\cite{DBLP:conf/cvpr/TanakaIHNS025,DBLP:conf/acl/AskariSGB25,DBLP:conf/acl/Dong0D0DW25}.
This capability enables MLLMs to serve not only as generators but also as integrative engines for multi-modal retrieval-augmented generation (mRAG).
By jointly processing retrieved multi-modal evidence and queries within a unified reasoning backbone, MLLMs can synthesize information and derive coherent conclusions from retrieved knowledge~\cite{DBLP:journals/corr/abs-2410-03577,DBLP:conf/eccv/ZhangCZLCYL24,DBLP:conf/cvpr/MaGS0LR025}.
Nonetheless, when confronted with complex and multi-faceted multi-modal queries that require diverse expertise or deliberative reasoning, a single agent often reaches its performance limit~\cite{DBLP:journals/corr/abs-2508-07023,DBLP:conf/miccai/ChenJYLWQZ24,DBLP:conf/acl/YangHLH25}.

\begin{figure*}[!t]
\centering
\includegraphics[width=15.5cm]{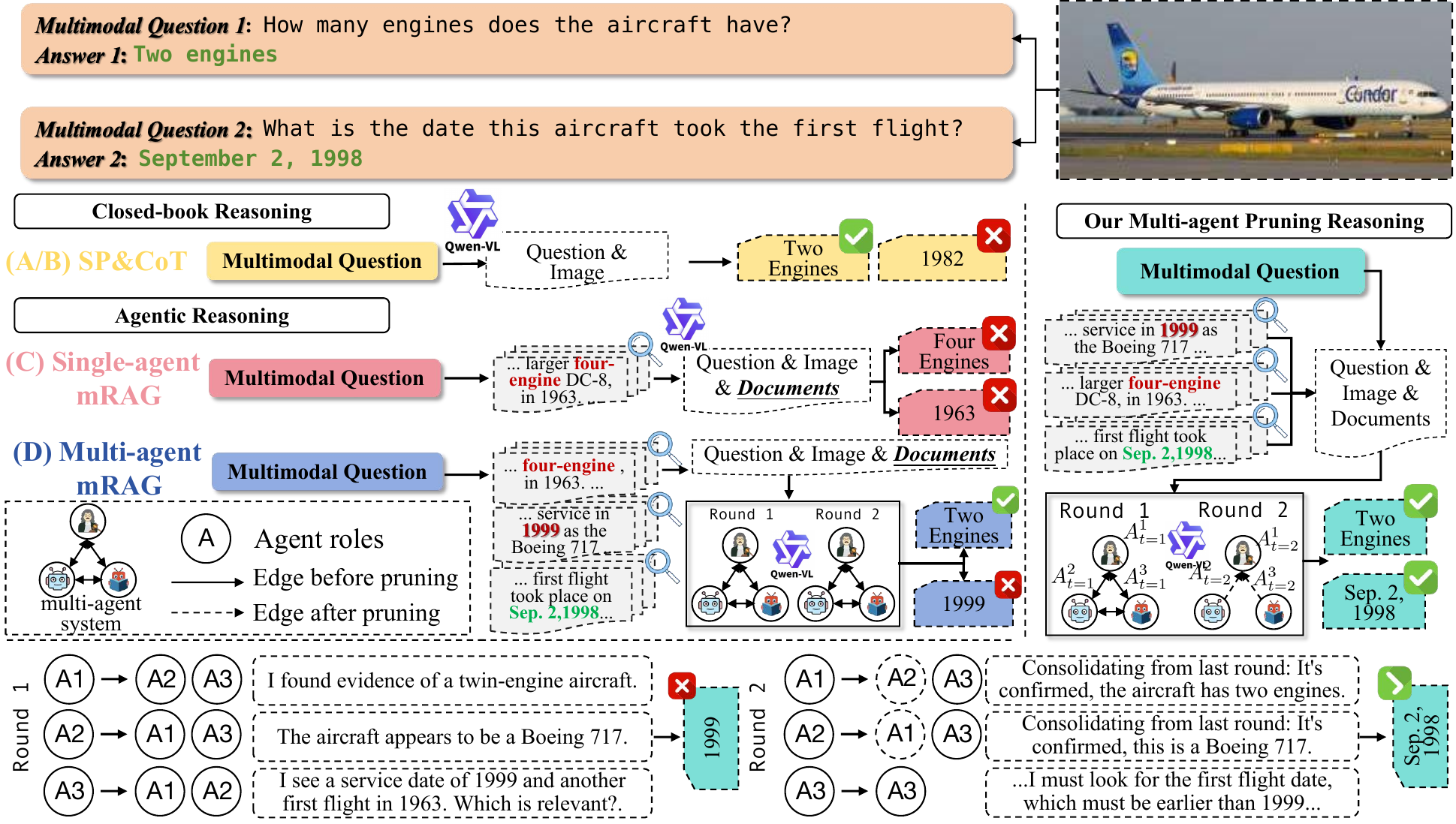}
\vspace{-.5em}
\caption{
Comparison of our approach with existing methods.
(1) \texttt{Closed-book Reasoning} does not utilize external knowledge.
(2) \texttt{Single-agent mRAG} leverages an end-to-end MLLM coupled with a retriever to answer all questions.
(3) \texttt{Multi-agent mRAG} constructs a fixed communication topology to obtain collaborative answers, without considering communication efficiency.
(4) Our \texttt{Multi-agent Pruning} method prunes unnecessary communication edges, improving answer quality and token efficiency.
}
\label{m3prune_motivation}
\vspace{-1.25em}
\end{figure*}

Recent studies have explored multi-agent systems built upon MLLMs, where multiple specialized agents communicate through structured graph topologies to distribute reasoning workloads for complex multi-modal tasks~\cite{DBLP:conf/cvpr/YueZ0D025,DBLP:conf/cvpr/PerincherryKL25,DBLP:conf/acl/MenJC00025}.
However, such performance gains often come at the cost of substantial computational and token overhead.
We identify the root cause as \texttt{communication redundancy}, which introduces inefficiency and can even undermine the accuracy of the final response~\cite{DBLP:conf/iclr/ZhangYLYWWCY025,DBLP:conf/naacl/JiangXWYHBSTM25}.
This redundancy is especially pronounced in multi-modal settings, where the information required to process textual and visual cues differs.
Yet existing methods frequently adopt uniform communication strategies across modalities~\cite{DBLP:conf/iclr/LiLWJZZWZH0Y25,DBLP:journals/corr/abs-2508-08816,DBLP:journals/corr/abs-2504-12330}.
As illustrated in Fig.~\ref{m3prune_motivation}, in the ``first flight date'' example, the date-expert agent is overwhelmed by irrelevant messages about engine counts and model variants.
Such noise hinders evidence aggregation and increases the likelihood of an incorrect answer.

In this paper, we introduce \textbf{M$^3$Prune}, a framework for hierarchical communication-graph pruning in multi-agent mRAG systems.
M$^3$Prune learns task-adaptive communication topologies by assigning learnable softened adjacency matrices to spatio-temporal intra- and inter-modal connections, optimizing them for task utility under sparsity regularization, and progressively pruning edges with low importance.
The core components of M$^3$Prune are:

\noindent\textbf{Intra-Modal Graph Sparsification:}
In mRAG tasks, agents assigned different roles may produce divergent responses to the same query~\cite{DBLP:conf/acl/ZhuDHYGWWQTJY25,DBLP:journals/corr/abs-2508-03404}.
Consequently, both cooperation and conflict may arise among agents within the same modality, potentially limiting mutual improvement.
To address this issue, we design a spatio-temporal message-passing scheme that facilitates effective response exchange within the visual and textual modalities, respectively.
We learn softened adjacency matrices that quantify the importance of intra-modal communication links based on their contribution to the final response.
Edges that contribute little are sparsified, thereby focusing communication on key agents.

\noindent\textbf{Inter-Modal Graph Sparsification:}
Given the heterogeneous granularity of information across modalities, it is crucial to integrate key clues from multiple modalities before generating the final response~\cite{DBLP:conf/iccv/CaiPYN023,DBLP:conf/eccv/LinWTWZWTMLSYYZ24,DBLP:conf/acl/WangSHZ25,DBLP:conf/acl/AbootorabiZDMMG25}.
Accordingly, we construct an inter-modal spatio-temporal communication graph to enable robust cross-modal collaboration and mitigate inconsistent responses.
We initialize inter-modal connections and learn their softened adjacency matrices to assess the importance of cross-modal communication.
In addition to optimizing task utility and sparsity regularization, we introduce a modality alignment score to encourage consistent semantic understanding across modalities during inter-modal sparsification.
Finally, after learning sparsified intra- and inter-modal graphs, we progressively prune redundant edges to obtain an efficient hierarchical topology.

In experiments, we evaluate M$^3$Prune against strong baselines in zero-shot, single-agent, and multi-agent settings on several mRAG benchmarks: Vidoseek~\cite{DBLP:journals/corr/abs-2502-18017}, MultimodalQA~\cite{talmor2021multimodalqa}, and ScienceQA~\cite{DBLP:conf/nips/LuMX0CZTCK22}, covering both general and domain-specific tasks.
The results show that M$^3$Prune achieves state-of-the-art performance, improving accuracy by \textbf{9.4\%} and token efficiency by \textbf{23.8\%} compared with strong multi-agent multi-modal baselines.

\section{Related Work}

\subsection{Multi-Modal RAG}
Prior research on mRAG can be broadly grouped into two themes:
(1) \textbf{Modality-Adaptive Retrieval Strategies.} Early mRAG systems typically employed static retrieval methods, often incurring unnecessary computational overhead~\cite{DBLP:conf/cvpr/ChunORKL21,DBLP:conf/nips/LuMX0CZTCK22,DBLP:conf/iccv/Luo0XGSTMLJ23}.
More recent work introduces dynamic, query-aware retrieval strategies that determine not only whether to retrieve but also which modality to retrieve from.
For example, EchoSight~\cite{DBLP:conf/emnlp/YanX24} and RoRA-VLM~\cite{DBLP:journals/corr/abs-2410-08876} perform wiki-article retrieval using visual-only information, followed by re-ranking with respect to the combined text--visual query.
(2) \textbf{Retrieval-Aware Pre-Training and Fine-Tuning.}
CoRe-MMRAG~\cite{DBLP:conf/acl/TianLZWHN25} addresses inconsistencies across knowledge sources through a four-stage progressive multi-modal retrieval pipeline.
LLaVA-mR$^2$AG~\cite{DBLP:journals/corr/abs-2411-15041} introduces a fine-tuned adaptive retriever that derives answers through two reflection operations.
Wiki-LLaVA~\cite{DBLP:conf/cvpr/CaffagniCMSC0C22} employs a hierarchical retrieval pipeline to integrate external knowledge from multi-modal documents.
While these methods rely on a single MLLM backbone, they largely overlook the potential benefits of collective intelligence and multi-agent collaboration for mRAG~\cite{DBLP:conf/naacl/JiangXWYHBSTM25}.

\subsection{Multi-Agent Systems for mRAG}
This line of work can be broadly categorized into the following directions:
(1) \textbf{General Agentic Collaboration.} A key shift has been the transition from static retrieval-and-generation pipelines to agentic designs, transforming passive systems into active decision-making frameworks~\cite{DBLP:journals/corr/abs-2508-03404,DBLP:journals/corr/abs-2509-24314,DBLP:conf/cvpr/Capellera0FA25}.
ViDoRAG~\cite{DBLP:journals/corr/abs-2502-18017} introduces an iterative agent workflow that combines exploration, summarization, and reflection to enable scalable test-time reasoning in mRAG.
OmniSearch~\cite{DBLP:conf/iclr/LiLWJZZWZH0Y25} designs a self-adaptive planning agent that emulates human behavior by dynamically decomposing complex multi-modal queries into sub-questions with adaptive retrieval steps.
E-agent~\cite{DBLP:journals/corr/abs-2508-08816} proposes a plan-then-execute architecture that combines a dynamic mRAG planner with a tool executor.
(2) \textbf{Evolving Collaborative Architectures.} As mRAG tasks grow more complex, general multi-agent systems encounter increasing limitations, motivating research on specialized multi-modal agent collaboration~\cite{DBLP:conf/acl/Dong0D0DW25,ranaldietal}.
HM-RAG~\cite{DBLP:journals/corr/abs-2504-12330} employs a hierarchical agent structure, including a decomposer agent to break down multi-intent queries and a decision agent to synthesize the final answer.
MuaLLM~\cite{DBLP:journals/corr/abs-2508-08137} integrates a hybrid mRAG framework with iterative reasoning, goal setting, and multi-step retrieval using an adaptive vector database.
Although multi-agent collaboration yields considerable improvements on complex mRAG tasks, redundant and inefficient agent interactions remain a bottleneck, increasing both computational and token costs~\cite{DBLP:conf/naacl/JiangXWYHBSTM25,DBLP:conf/iclr/ZhangYLYWWCY025}.
To address this issue, we propose a multi-modal multi-agent edge-pruning framework that preserves task performance while reducing token overhead.

\section{Basic Notations and Task Definition}

\noindent\textbf{Basic Notations.}
Our framework consists of intra- and inter-modal graph pruning modules.\footnote{All notations and agent prompt descriptions are summarized in Appendix~\ref{prompt_des}.}
We represent each modality-specific communication structure as an augmented graph that encodes both connectivity and agent memory states.
The intra-modal module comprises a textual graph,
$\mathcal{G}^{\text{intra}}_{txt} = (\mathcal{V}_{txt}, \mathcal{E}^\mathcal{T}_{txt}, \mathcal{E}^\mathcal{S}_{txt}, \mathcal{S}_{txt})$,
and a visual graph,
$\mathcal{G}^{\text{intra}}_{vis} = (\mathcal{V}_{vis}, \mathcal{E}^\mathcal{T}_{vis}, \mathcal{E}^\mathcal{S}_{vis}, \mathcal{S}_{vis})$,
where $\mathcal{V}_{txt} = \{v^{txt}_i\}$ and $\mathcal{V}_{vis} = \{v^{vis}_i\}$ denote the sets of textual and visual agent nodes, respectively.
Each agent node is initialized with a corresponding textual or visual role.
$\mathcal{E}^\mathcal{T}_{txt}$ and $\mathcal{E}^\mathcal{S}_{txt}$ denote the textual temporal and spatial edge sets, respectively,\footnote{Each agent is fully connected to all other agents except itself through spatial edges, and connected to all agents (including itself) through temporal edges across rounds.}
and $\mathcal{S}_{txt}^{(t)} = \{ s^{txt,(t)}_i\}$ represents the memory states of textual agents at round $t$.
Correspondingly, $\mathcal{E}^\mathcal{T}_{vis}$ and $\mathcal{E}^\mathcal{S}_{vis}$ denote the visual temporal and spatial edge sets, and $\mathcal{S}_{vis}^{(t)} = \{ s^{vis,(t)}_i \}$ represents the memory states of visual agents at round $t$.
The memory update of each agent from round $t$ to round $(t+1)$ is:
\begin{equation}
\label{eq_1}
s_{i}^{m,(t+1)} = f_{\text{tr}} \bigl(s_{i}^{m,(t)}, \mathbf{q}, \mathbf{c}, I_{\mathcal{T}}^{m,(t+1)}, I_{\mathcal{S}}^{m,(t+1)} \bigr),
\end{equation}
where $f_{\text{tr}}$ denotes the MLLM-based aggregation function that updates the agent memory at the current round.
Here, $m \in \{\text{txt}, \text{vis}\}$ indexes the modality, and $\mathbf{q}$ and $\mathbf{c}$ denote the question and retrieved contexts, respectively.
$I_{\mathcal{T}}^{m,(t+1)}$ and $I_{\mathcal{S}}^{m,(t+1)}$ denote the information aggregated from temporal and spatial neighbors, respectively, in modality $m$ at round $(t+1)$.

For the inter-modal component, all textual and visual agents are incorporated into an inter-modal graph,
$\mathcal{G}^{\text{inter}} = (\mathcal{V}, \mathcal{E}^\mathcal{T}, \mathcal{E}^\mathcal{S}, \mathcal{S})$, where the node set is defined as $\mathcal{V} = \mathcal{V}_{txt} \cup \mathcal{V}_{vis}$.
The temporal edge set integrates both intra- and inter-modal connections:
$\mathcal{E}^\mathcal{T} = \mathcal{E}^\mathcal{T}_{txt} \cup \mathcal{E}^\mathcal{T}_{vis} \cup \mathcal{E}^\mathcal{T}_{vis \rightarrow txt} \cup \mathcal{E}^\mathcal{T}_{txt \rightarrow vis}$.
Similarly, the spatial edge set is defined as:
$\mathcal{E}^\mathcal{S} = \mathcal{E}^\mathcal{S}_{txt} \cup \mathcal{E}^\mathcal{S}_{vis} \cup \mathcal{E}^\mathcal{S}_{vis \rightarrow txt} \cup \mathcal{E}^\mathcal{S}_{txt \rightarrow vis}$.
The intra-modal edges in both temporal and spatial sets are inherited from the corresponding intra-modal graphs, while the inter-modal edges are initialized as fully connected.
Finally, the joint agent memory is denoted as $\mathcal{S} = \mathcal{S}_{txt} \cup \mathcal{S}_{vis}$.
The memory update mechanism is analogous to Eq.~\ref{eq_1}, except that the aggregated information $I_{\mathcal{T}}^{(t+1)}$ and $I_{\mathcal{S}}^{(t+1)}$ may also originate from inter-modal temporal and spatial edges. For example, $I_{\mathcal{T}}^{(t+1)}$ aggregates information from $I_{\mathcal{T}}^{txt,(t+1)}$ and $I_{\mathcal{T}}^{vis,(t+1)}$.

\noindent\textbf{Task Definition.}
Given a question $\mathbf{q}$ and retrieved contexts $\mathbf{c}$, each agent in an intra-modal graph receives information solely through edges within the same modality, whereas agents in the inter-modal graph may also receive information from agents in the other modality.
For each agent, its output response at round $t$ is defined as
$\mathcal{O}_{i}^{(t)} = f_\theta \left(\mathbf{q}, \mathbf{c}, I_{\mathcal{T}}^{(t)}, I_{\mathcal{S}}^{(t)} \right)$,
where $f_\theta$ denotes the MLLM-generated response conditioned on the question, retrieved contexts, and both temporal and spatial information available to the agent.
Here, $I_{\mathcal{T}}^{(t)}$ and $I_{\mathcal{S}}^{(t)}$ refer to the temporal and spatial information aggregated for the current agent at round $t$ from all available neighbors under the chosen communication graph (intra-modal or inter-modal).
After multiple rounds of multi-agent interaction, a summary agent produces the final answer:
$\mathcal{O}_\text{s}^{(T)} = f_\text{s} \left(\mathbf{q}, \mathbf{c}, I_{\mathcal{T}}^{(T)}, I_{\mathcal{S}}^{(T)} \right)$,
where $T$ is the total number of rounds, $f_\text{s}$ denotes the summary agent, and $\mathcal{O}_\text{s}^{(T)}$ is the final answer.

However, a fixed communication topology inevitably introduces redundant information, motivating the need to prune superfluous edges~\cite{DBLP:conf/iclr/LiLWJZZWZH0Y25,DBLP:journals/corr/abs-2508-08816}.
To formalize this issue, we define communication redundancy to guide the pruning process:

\noindent\textbf{Definition 1 (Communication Redundancy).}
Given a multi-modal multi-agent communication graph
$\mathcal{G} = (\mathcal{V}, \mathcal{E}^\mathcal{T}, \mathcal{E}^\mathcal{S}, \mathcal{S})$
with a utility function $\phi(\cdot)$ that quantifies task performance, an edge $e \in \mathcal{E}^\mathcal{T} \cup \mathcal{E}^\mathcal{S}$ is considered redundant if removing it does not degrade task performance, i.e.,
\begin{equation}
\label{communication_redundancy_eq}
\mathcal{G}^{sub} = (\mathcal{V}, \tilde{\mathcal{E}}^\mathcal{T}, \tilde{\mathcal{E}}^\mathcal{S}, \mathcal{S}), \quad
\phi(\mathcal{G}^{sub}) \geq \phi(\mathcal{G}),
\end{equation}
where
$\tilde{\mathcal{E}}^\mathcal{T} =
\begin{cases}
\mathcal{E}^\mathcal{T} \setminus \{e\}, & e \in \mathcal{E}^\mathcal{T}, \\
\mathcal{E}^\mathcal{T}, & e \in \mathcal{E}^\mathcal{S},
\end{cases}$
and
$\tilde{\mathcal{E}}^\mathcal{S} =
\begin{cases}
\mathcal{E}^\mathcal{S} \setminus \{e\}, & e \in \mathcal{E}^\mathcal{S}, \\
\mathcal{E}^\mathcal{S}, & e \in \mathcal{E}^\mathcal{T}.
\end{cases}$

Specifically, the goal of inter-modal edge pruning is to construct a subgraph $\mathcal{G}^{sub}$ by removing redundant edges while preserving task performance~\cite{DBLP:conf/iclr/ZhangYLYWWCY025}; this objective is further regularized by modality semantic alignment~\cite{DBLP:conf/acl/AbootorabiZDMMG25}, as elaborated next.

\begin{figure*}[!t]
\centering
\includegraphics[width=17cm]{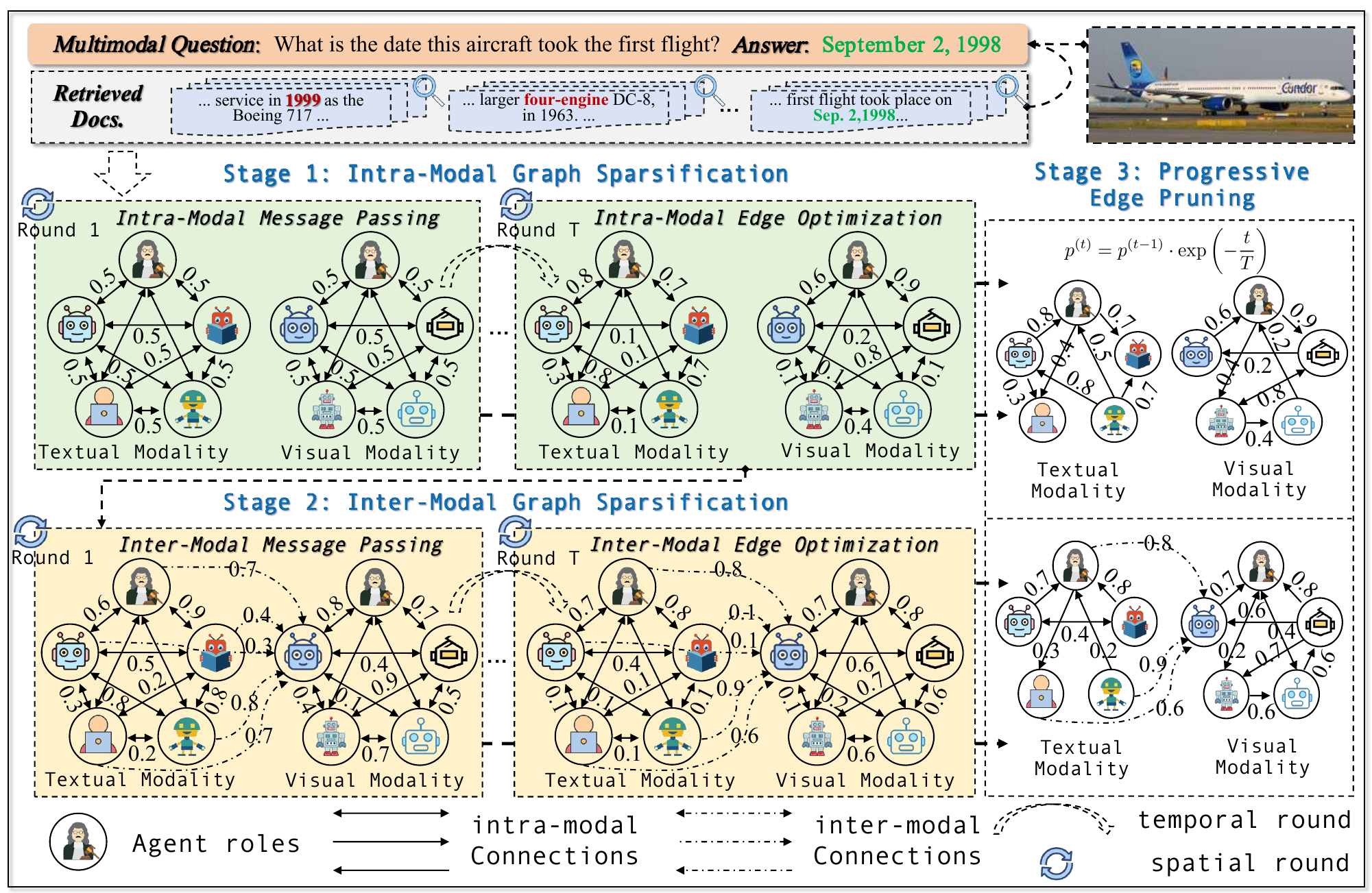}
\vspace{-0.2em}
\caption{
Overview. The key components include:
(1) \texttt{Intra-modal Graph Sparsification}, which analyzes the input question using multiple agents for the textual and visual modalities, respectively;
(2) \texttt{Inter-modal Graph Sparsification}, which integrates semantic information across modalities through interactions among multi-agent responses in the textual and visual domains;
(3) \texttt{Progressive Edge Pruning}, which progressively prunes redundant edges \emph{during training} to obtain a compact communication graph for inference.
Due to the large number of connections in the inter-modal stage, we illustrate only the interactions of one agent (dashed lines) as an example.
}
\label{model_fig}
\vspace{-.5em}
\end{figure*}
\section{Methodology}

\subsection{Intra-Modal Graph Sparsification}
\label{intra_modal_sec}
We design a spatio-temporal message-passing scheme to facilitate the effective exchange of task-relevant responses among agents within the visual and textual modalities.
To improve the quality of the final response, we dynamically adjust edge weights and sparsify redundant connections, thereby focusing communication on critical interactions.

\noindent\textbf{Edge Weight Initialization.}
Given the predefined spatial and temporal edges $\mathcal{E}_{m}^\mathcal{S}$ and $\mathcal{E}_{m}^\mathcal{T}$ for $m \in \{\text{txt}, \text{vis}\}$, we obtain the corresponding adjacency matrices $\mathbf{A}^\mathcal{S}_{m}$ and $\mathbf{A}^\mathcal{T}_{m}$ for the intra-modal communication graph.
Based on $\mathbf{A}^\mathcal{S}_{m}$ and $\mathbf{A}^\mathcal{T}_{m}$, we initialize learnable softened adjacency matrices $\tilde{\mathbf{A}}^\mathcal{S}_{m}$ and $\tilde{\mathbf{A}}^\mathcal{T}_{m}$ as edge weights via Gumbel-Softmax~\cite{jang2016categorical} (Eq.~\ref{eq_6}), yielding values in $(0,1)$.
Taking $\tilde{\mathbf{A}}^\mathcal{S}_{m}$ as an example, each element $\tilde{\mathbf{A}}^\mathcal{S}_{m}[i,j]$ is computed as:
\begin{equation}
\label{eq_6}
\tilde{\mathbf{A}}^\mathcal{S}_{m}[i,j] = \frac{\exp\left((\log({\mathbf{A}}^\mathcal{S}_{m}[i,j] + \epsilon) + g_{ij})/\tau\right)}{\sum_{k=1}^{N_m} \exp\left((\log({\mathbf{A}}^\mathcal{S}_{m}[i,k] + \epsilon) + g_{ik})/\tau\right)}
\end{equation}
where $g_{ij} \sim \text{Gumbel}(0,1)$, $\tau$ is the temperature, $N_{m}$ is the number of agents in modality $m$, and $\epsilon$ is a small constant for numerical stability.
$\tilde{\mathbf{A}}^\mathcal{T}_{m}$ is obtained similarly.
To enable multi-round communication, we maintain round-specific softened adjacency matrices $\tilde{\mathbf{A}}_{m}^{(\mathcal{S},(t))}$ and $\tilde{\mathbf{A}}_{m}^{(\mathcal{T},(t))}$, which are updated at each round $t$.

\noindent\textbf{Intra-Modal Message Passing.}
Within the intra-modal graph $\mathcal{G}^\text{intra}$, agent $v^{m,(t)}_i$ at round $t$ receives the spatial message:
\begin{equation}
\mathcal{M}_{i,\text{intra}}^{(m,\mathcal{S},(t))} = \sum_{v_j^{m,(t)} \in \mathcal{N}_\text{intra}^\mathcal{S}(v_i^{m,(t)})} \mathcal{W}_{m}^{(\mathcal{S},(t))}[i,j] \cdot \mathcal{O}(v_j^{m,(t)}),
\end{equation}
where the edge attention weights are defined as
\begin{equation}
\label{eq_8}
\mathcal{W}_{m}^{(\mathcal{S},(t))}[i,j] = \frac{\exp(\tilde{\mathbf{A}}_{m}^{(\mathcal{S},(t))}[i,j])}{\sum_{v_k^{m,(t)} \in \mathcal{N}_\text{intra}^\mathcal{S}(v_i^{m,(t)})} \exp(\tilde{\mathbf{A}}_{m}^{(\mathcal{S},(t))}[i,k])}.
\end{equation}
Here, $\mathcal{N}_\text{intra}^\mathcal{S}(v_i^{m,(t)})$ denotes the set of spatial neighbors of $v_i^{m,(t)}$, and $\mathcal{O}(v_j^{m,(t)})$ denotes the output response of agent $v_j^{m,(t)}$ at round $t$.
Since an agent may be connected to multiple neighbors that contribute unequally, Eq.~\ref{eq_8} normalizes the aggregation weights accordingly.
This procedure applies analogously to temporal message passing.
After aggregating spatio-temporal information, each node leverages the MLLM to update its response.
After $T$ rounds, the summary agent synthesizes messages from the two modalities to produce the final answer.

\noindent\textbf{Intra-Modal Edge Optimization.}
The edge optimization objective maximizes the expected utility while penalizing the nuclear norm of the softened adjacency matrices, thereby balancing task performance and communication-graph sparsity.
Accordingly, the training objective is defined as follows:
\begin{equation}
\max_{\tilde{\mathbf{A}}_{m}^\mathcal{S},\,\tilde{\mathbf{A}}_{m}^\mathcal{T}} \mathbb{E}_{{\mathcal{G}}^\text{intra}}\left[\phi({\mathcal{G}}^\text{intra}) \right] - \|\tilde{\mathbf{A}}_{m}^\mathcal{S}\|_{*} - \|\tilde{\mathbf{A}}_{m}^\mathcal{T}\|_{*}
\end{equation}
subject to $\| \mathbf{A}_{m}^\mathcal{S} - \tilde{\mathbf{A}}_{m}^\mathcal{S} \|_F + \| \mathbf{A}_{m}^\mathcal{T} - \tilde{\mathbf{A}}_{m}^\mathcal{T} \|_F \le \delta$,
where $\delta$ defines the noise tolerance level, $\phi(\cdot)$ denotes the task performance metric, and $\|\cdot\|_{*}$ denotes the nuclear norm.
Because the optimization process may involve API calls to the underlying MLLM, which are non-differentiable, we adopt the policy gradient method~\cite{williams1992simple} to approximate gradients.
Following~\cite{DBLP:conf/iclr/ZhangYLYWWCY025}, we optimize the objective by sampling $K$ graph instances from $\mathcal{G}^\text{intra}$, denoted by ${\mathcal{G}}^\text{intra}_k$. The sampling and optimization process is formulated as follows:
\begin{equation}
\label{eq_pg}
\nabla \mathbb{E}_{{\mathcal{G}}^\text{intra}}[\phi({\mathcal{G}}^\text{intra})] \approx {}
\frac{1}{K}\sum_{k=1}^K \phi({\mathcal{G}}^\text{intra}_k) \cdot \nabla \log \mathcal{P}({\mathcal{G}}^\text{intra}_k)
\end{equation}
\begin{equation}
    \mathcal{P}({\mathcal{G}}^\text{intra}_k) =
\left(\prod_{t=1}^T \prod_{e_{ij} \in \mathcal{E}_{m}^{\mathcal{S},k,(t)}} \tilde{\mathbf{A}}_{m}^{\mathcal{S},(t)}[i,j] \right)
\cdot \left(\prod_{t=2}^T \prod_{e_{ij} \in \mathcal{E}_{m}^{\mathcal{T},k,(t)}} \tilde{\mathbf{A}}_{m}^{\mathcal{T},(t)}[i,j] \right)
\end{equation}
where $\mathcal{P}({\mathcal{G}}^\text{intra}_k)$ is the sampling probability of ${\mathcal{G}}^\text{intra}_k$, and $K$ is the number of sampled graphs.

\begin{table*}[!t]
    \scriptsize
    \centering
    \caption{Performance comparison between M$^3$Prune and baselines on the domain-specific ScienceQA benchmark. The t-tests demonstrate the improvements of our work are statistically significant with $p < 0.05$ level.}
    \vspace{-1em}
    \setlength{\tabcolsep}{2.83pt}
    \renewcommand{\arraystretch}{0.85}
    \begin{tabular}{cc|ccc|ccc|cc|c|ccc|ccc|cc|c}
        \toprule
        \multirow{5}*[-0.85em]{\makecell{\textbf{Training} \\ \textbf{Paradigm}}}
        & \multirow{5}*[-0.85em]{\textbf{Method}} & \multicolumn{9}{c|}{\raisebox{-0.5em}{\textbf{Llama3.2-VL (11B)}}} & \multicolumn{9}{c}{\raisebox{-0.5em}{\textbf{Qwen-VL-Max}}}  \\
        & & \multicolumn{9}{c|}{\rule{204pt}{0.4pt}} & \multicolumn{9}{c}{\rule{204pt}{0.4pt}} \\
        & & \multicolumn{3}{c}{\raisebox{-0.5em}{\textbf{Subject}}} & \multicolumn{3}{c}{\raisebox{-0.5em}{\textbf{Context Modality}}} & \multicolumn{2}{c|}{\raisebox{-0.5em}{\textbf{Grade}}} & \multirow{3}{*}{\makecell{\textbf{Avg.}}} & \multicolumn{3}{c}{\raisebox{-0.5em}{\textbf{Subject}}} & \multicolumn{3}{c}{\raisebox{-0.5em}{\textbf{Context Modality}}} & \multicolumn{2}{c|}{\raisebox{-0.5em}{\textbf{Grade}}} & \multirow{3}{*}{\makecell{\textbf{Avg.}}} \\
        & & \multicolumn{3}{c}{\rule{60pt}{0.4pt}} & \multicolumn{3}{c}{\rule{60pt}{0.4pt}} & \multicolumn{2}{c|}{\rule{40pt}{0.4pt}} & & \multicolumn{3}{c}{\rule{60pt}{0.4pt}} & \multicolumn{3}{c}{\rule{60pt}{0.4pt}} & \multicolumn{2}{c|}{\rule{40pt}{0.4pt}} \\ 
        & & \textbf{NAT} & \textbf{Soc} & \textbf{LAN} & \textbf{TXT} & \textbf{IMG} & \textbf{NO} & \textbf{G1-6} & \textbf{G7-12} & & \textbf{NAT} & \textbf{Soc} & \textbf{LAN} & \textbf{TXT} & \textbf{IMG} & \textbf{NO} & \textbf{G1-6} & \textbf{G7-12}  \\
        \midrule

        \multirow{2}{*}{Zero-shot} & SP & 82.92 & 88.79 & 77.55 & 82.87 & 82.47 & 79.09 & 84.30 & 79.99 & 82.76$_{\pm 1.2}$ & 91.90 & 90.56 & 89.36 & 91.47 & 88.09 & 90.99 & 91.44 & 90.10 & 90.96$_{\pm 0.9}$  \\
        & CoT & 84.89 & \textbf{97.95} & 69.00 & \underline{85.88} & \textbf{93.70} & 70.51 & 87.36 & 76.59 & 83.51$_{\pm 0.5}$ & 92.36 & 91.56 & 90.45 & 91.64 & 88.99 & 91.99 & 92.03 & 91.10 & 91.70$_{\pm 0.8}$  \\
        \cmidrule(lr){1-20}
        \multirow{5}{*}{\makecell{Single-agent \\ RAG}} & Wiki-LLaVA & 82.30 & 92.21 & 80.25 & 83.69 & 84.66 & 80.21 & 84.80 & 82.17 & 83.85$_{\pm 0.8}$ & 91.00 & 95.62 & 89.98 & 91.84 & 91.45 & 89.44 & 93.13 & 89.10 & 91.69$_{\pm 0.4}$ \\
        & RoRA-VLM & 82.23 & 92.11 & 80.20 & 83.64 & 84.61 & 80.13 & 84.77 & 82.13 & 83.80$_{\pm 0.9}$ & 90.92 & 95.51 & 89.88 & 91.78 & 91.40 & 89.39 & 93.04 & 89.07 & 91.65$_{\pm 1.5}$ \\
        & EchoSight & 82.35 & 92.23 & 80.27 & 83.70 & 84.69 & 80.23 & 84.81 & 82.21 & 83.88$_{\pm 1.1}$ & 91.04 & 95.65 & 90.00 & 91.87 & 91.47 & 89.48 & 93.19 & 89.12 & 91.73$_{\pm 0.8}$  \\
        & LLaVA-mR2AG & 82.37 & 92.22 & 80.31 & 83.73 & 84.71 & 80.26 & 84.85 & 82.23 & 83.90$_{\pm 0.8}$ & 91.07 & 95.68 & 90.02 & 91.90 & 91.48 & 89.51 & 93.22 & 89.17 & 91.80$_{\pm 0.6}$ \\
        & CoRe-MMRAG & 82.44 & 92.31 & 80.30 & 83.77 & 84.73 & 80.28 & 84.86 & \underline{82.27} & 83.94$_{\pm 1.2}$ & 91.10 & \underline{95.72} & 90.06 & 91.93 & 91.50 & 89.55 & 93.24 & 89.18 & 91.82$_{\pm 1.0}$ \\
        \cmidrule(lr){1-20}
        \multirow{5}{*}{\makecell{Multi-agent \\ RAG}} & OmniSearch & 82.71 & 88.13 & \textbf{83.36} & 82.61 & 80.56 & 83.33 & 85.25 & 81.80 & 84.01$_{\pm 0.3}$ & 92.50 & 93.43 & 90.09 & 93.35 & 91.31 & 89.71 & 93.35 & 89.76 & 92.07$_{\pm 0.4}$ \\
        & ViDoRAG & 83.88 & 92.15 & 83.00 & 83.97 & 84.96 & 82.39 & 87.43 & 81.72 & \underline{85.39}$_{\pm 0.7}$ & 93.99 & 92.20 & 90.45 & 93.91 & 90.26 & 91.29 & 93.79 & 90.74 & 92.70$_{\pm 0.5}$ \\
        & HM-RAG & \textbf{86.25} & \underline{94.89} & 72.45 & \textbf{86.56} & \underline{89.70} & 75.23 & \underline{87.78} & 79.45 & 84.80$_{\pm 1.3}$ & \underline{94.30} & 93.66 & \underline{92.36} & 93.47 & 90.36 & \underline{94.08} & \underline{94.15} & \underline{92.78} & \underline{93.66}$_{\pm 1.2}$  \\
        & E-Agent & 83.00 & 87.58 & \underline{83.09} & 82.45 & 80.40 & \underline{83.51} & 85.31 & 81.59 & 83.98$_{\pm 1.4}$ & 93.67 & 93.44 & 89.18 & \underline{94.06} & \underline{91.84} & 89.90 & 93.20 & 91.13 & 92.46$_{\pm 1.5}$ \\
       \rowcolor[HTML]{C0C0C0} & \textbf{Ours} & \underline{85.97} & 94.60 & 81.45 & 85.29 & 86.47 & \textbf{83.83} & \textbf{88.25} & \textbf{83.65} & \textbf{86.61}$_{\pm 0.6}$ & \textbf{97.51} & \textbf{96.63} & \textbf{97.82} & \textbf{97.70} & \textbf{96.13} & \textbf{97.98} & \textbf{98.05} & \textbf{96.24} & \textbf{97.41}$_{\pm 0.4}$ \\
        \bottomrule
    \end{tabular}
    \label{scienceQA}
\end{table*}

\subsection{Inter-Modal Graph Sparsification}
\label{inter_modal_sec}
Building on the intra-modal graphs, we integrate cross-modal information and learn the importance of cross-modal edges.
The inter-modal topology is optimized for both task performance and structural regularity.
Additionally, we introduce a modality alignment score to encourage consistent task understanding across modalities during sparsification.

\noindent\textbf{Edge Weight Initialization.}
Recall that the intra-modal edges ($\mathcal{E}_{m}^\mathcal{S}$, $\mathcal{E}_{m}^\mathcal{T}$), adjacency matrices ($\mathbf{A}^\mathcal{S}_{m}$, $\mathbf{A}^\mathcal{T}_{m}$), and softened adjacency matrices ($\tilde{\mathbf{A}}^\mathcal{S}_{m}$, $\tilde{\mathbf{A}}^\mathcal{T}_{m}$) are inherited from the preceding intra-modal phase.
Given predefined inter-modal spatial edges $\mathcal{E}_{m'}^\mathcal{S}$ and temporal edges $\mathcal{E}_{m'}^\mathcal{T}$, each agent is connected to all agents in the other modality, where $m' \in \{\text{txt} \rightarrow \text{vis},\, \text{vis} \rightarrow \text{txt}\}$ denotes the two inter-modal directions.
We obtain the initial inter-modal adjacency matrices $\mathbf{A}^\mathcal{S}_{m'}$ and $\mathbf{A}^\mathcal{T}_{m'}$.
Similarly, we initialize the softened adjacency matrices of the inter-modal edges,
$\tilde{\mathbf{A}}^{\mathcal{S}}_{m'}$ and $\tilde{\mathbf{A}}^{\mathcal{T}}_{m'}$, following Eq.~\ref{eq_6}.

\noindent\textbf{Inter-Modal Message Passing.}
Each agent $v_i^{(t)}$ at round $t$ receives neighboring information from both modalities through $\mathcal{G}^\text{inter}$.
We first consider spatial message passing:
\begin{equation}
\mathcal{M}_{i,\text{inter}}^{\mathcal{S},(t)} = \sum_{v_j^{(t)} \in \mathcal{N}^\mathcal{S}_\text{inter}(v_i^{(t)})} \mathcal{W}^{(\mathcal{S},(t))}[i,j] \cdot \mathcal{O}(v_j^{(t)}),
\end{equation}
\begin{equation}
\label{eq_14}
\mathcal{W}^{(\mathcal{S},(t))}[i,j] = \frac{\exp(\tilde{\mathbf{A}}^{\mathcal{S}}_{m'}[i,j])}{\sum_{v_k^{(t)} \in \mathcal{N}_\text{inter}^\mathcal{S}(v_i^{(t)})} \exp(\tilde{\mathbf{A}}^{\mathcal{S}}_{m'}[i,k])},
\end{equation}
where $\mathcal{N}^\mathcal{S}_\text{inter}(v_i^{(t)})$ denotes the set of spatial neighbors of $v_i^{(t)}$ at round $t$.
Aggregation weights are computed as in Eq.~\ref{eq_14}, analogous to Eq.~\ref{eq_8}.
Temporal message passing is defined analogously.
Each node aggregates spatio-temporal information from both modality directions.
In addition, it incorporates the results of the intra-modal message passing discussed previously.
Finally, the agent generates its output response based on the MLLM backbone.
After $T$ rounds of multi-agent discussion, the summary agent synthesizes messages from all agents to generate the final answer.

\noindent\textbf{Inter-Modal Edge Optimization.}
Similar to intra-modal edge optimization, our training objective jointly optimizes task performance and graph sparsity.
We additionally introduce a modality alignment score to encourage consistent semantic understanding across modalities by preserving inter-modal associations:
\begin{align}
\max_{\tilde{\mathbf{A}}_\text{inter}} \ & \mathbb{E}_{\mathcal{G}^\text{inter}}
[\phi(\mathcal{G}^\text{inter})]
- \sum_{\mathcal{X} \in \{\mathcal{S},\mathcal{T}\}}\sum_{m' \in \{\text{txt} \rightarrow \text{vis},\, \text{vis} \rightarrow \text{txt}\}}\|\tilde{\mathbf{A}}^{\mathcal{X}}_{m'}\|_{*} \nonumber\\
& + \sum_{\mathcal{X} \in \{\mathcal{S},\mathcal{T}\}} \mathcal{L}_{\mathrm{align}}(\tilde{\mathbf{A}}^\mathcal{X}_{\text{txt} \rightarrow \text{vis}}, \tilde{\mathbf{A}}^\mathcal{X}_{\text{vis} \rightarrow \text{txt}})
\end{align}
subject to $\sum_{\mathcal{X} \in \{\mathcal{S},\mathcal{T}\}} \| \mathbf{A}_{\text{txt} \rightarrow \text{vis}}^\mathcal{X} - \tilde{\mathbf{A}}_{\text{txt} \rightarrow \text{vis}}^\mathcal{X} \|_F \le \delta$ and $\sum_{\mathcal{X} \in \{\mathcal{S},\mathcal{T}\}} \| \mathbf{A}_{\text{vis} \rightarrow \text{txt}}^\mathcal{X} - \tilde{\mathbf{A}}_{\text{vis} \rightarrow \text{txt}}^\mathcal{X} \|_F \le \delta$,
where $\delta$ is the noise tolerance level.
$\tilde{\mathbf{A}}_\text{inter}$ denotes the collection of four softened adjacency matrices, namely $\tilde{\mathbf{A}}^{\mathcal{S}}_{m'}$ and $\tilde{\mathbf{A}}^{\mathcal{T}}_{m'}$ with $m' \in \{\text{txt} \rightarrow \text{vis},\, \text{vis} \rightarrow \text{txt}\}$.
The modality alignment score is defined as:
\begin{equation}
\mathcal{L}_{\mathrm{align}}(\tilde{\mathbf{A}}^\mathcal{X}_{\text{txt} \rightarrow \text{vis}}, \tilde{\mathbf{A}}^\mathcal{X}_{\text{vis} \rightarrow \text{txt}}) =
\frac{1}{N_T N_I} \sum_{i}\frac{\tilde{\mathbf{A}}^\mathcal{X}_{\text{txt} \rightarrow \text{vis}}[i,:] \cdot \left(\tilde{\mathbf{A}}^\mathcal{X}_{\text{vis} \rightarrow \text{txt}}\right)^{\top}[i,:]}{\|\tilde{\mathbf{A}}^\mathcal{X}_{\text{txt} \rightarrow \text{vis}}[i,:]\| \cdot \|\left(\tilde{\mathbf{A}}^\mathcal{X}_{\text{vis} \rightarrow \text{txt}}\right)^{\top}[i,:]\|},
\end{equation}
where $N_T$ and $N_I$ denote the numbers of textual and visual agents, respectively, and $i$ indexes the rows of the corresponding matrices.

\begin{table*}[!t]
    \vspace{-0.2cm}
    \scriptsize
    \centering
    \caption{Performance comparison on general-domain benchmarks. Due to space limitation, we present the performance on Qwen2.5-VL (7B) in Appendix \ref{general_perf}. The t-tests demonstrate the improvements are statistically significant with $p < 0.05$ level. }
    \vspace{-1em}
    \setlength{\tabcolsep}{1.8pt}
    \renewcommand{\arraystretch}{0.85}
    \begin{tabular}{ccc|cccccccccccc|cccc|cc}
        \toprule
        \multirow{5}*[-0.85em]{\textbf{Backbone}} &  \multirow{5}*[-0.85em]{\makecell{\textbf{Training} \\ \textbf{Paradigm}}}
        & \multirow{5}*[-0.85em]{\textbf{Method}} & \multicolumn{12}{c|}{\raisebox{-0.5em}{\textbf{Vidoseek}}} & \multicolumn{4}{c|}{\raisebox{-0.5em}{\textbf{MultimodalQA}}} & \multicolumn{2}{c}{\multirow{4}{*}{\textbf{Average}}} \\
        & & & \multicolumn{12}{c|}{\rule{226pt}{0.4pt}}  & \multicolumn{4}{c|}{\rule{70pt}{0.4pt}} \\
        &  & & \multicolumn{2}{c|}{\raisebox{-0.5em}{\textbf{Single-hop}}} & \multicolumn{2}{c|}{\raisebox{-0.5em}{\textbf{Multi-hop}}} & \multicolumn{2}{c|}{\raisebox{-0.5em}{\textbf{Text}}} & \multicolumn{2}{c|}{\raisebox{-0.5em}{\textbf{Table}}} & \multicolumn{2}{c|}{\raisebox{-0.5em}{\textbf{Chart}}} & \multicolumn{2}{c|}{\raisebox{-0.5em}{\textbf{Layout}}} &  \multicolumn{2}{c|}{\raisebox{-0.5em}{\textbf{Image}}} & \multicolumn{2}{c|}{\raisebox{-0.5em}{\textbf{Text}}} \\
       & & & \multicolumn{2}{c}{\rule{33pt}{0.4pt}} & \multicolumn{2}{c}{\rule{33pt}{0.4pt}} & \multicolumn{2}{c}{\rule{33pt}{0.4pt}} & \multicolumn{2}{c}{\rule{33pt}{0.4pt}} & \multicolumn{2}{c}{\rule{33pt}{0.4pt}} & \multicolumn{2}{c|}{\rule{33pt}{0.4pt}} & \multicolumn{2}{c}{\rule{33pt}{0.4pt}} & \multicolumn{2}{c|}{\rule{33pt}{0.4pt}} & \multicolumn{2}{c}{\rule{71pt}{0.4pt}} \\
       &  & & \textbf{Acc$^\star$} & \textbf{EM} & \textbf{Acc$^\star$} & \textbf{EM} & \textbf{Acc$^\star$} & \textbf{EM} & \textbf{Acc$^\star$} & \textbf{EM} & \textbf{Acc$^\star$} & \textbf{EM} & \textbf{Acc$^\star$} & \textbf{EM} & \textbf{Acc$^\star$} & \textbf{EM} & \textbf{Acc$^\star$} & \textbf{EM} & \textbf{Acc$^\star$} & \textbf{EM} \\
       \midrule
        
        \multirow{12}{*}{\makecell{Llama3.2-VL \\ (11B)}} & \multirow{2}{*}{Zero-shot} & SP & 25.89 & 2.48 & 12.07 & 8.45 & 26.25 & 0.01 & 9.14 & 5.14 & 14.65 & 10.83 & 22.88 & 4.38 & 14.68 & 15.23 & 36.28 & 31.46 & 20.23$_{\pm 0.8}$ & 9.75$_{\pm 0.6}$ \\
        & & CoT & 30.61 & 5.17 & 7.44 & 9.24 & 21.00 & 4.25 & 10.14 & 5.86 & 9.92 & 13.83 & 25.25 & 6.01 & 16.82 & 15.68 & 33.42 & 32.53 & 19.33$_{\pm 1.3}$ & 11.57$_{\pm 1.1}$ \\
        \cmidrule(lr){2-21}
        & \multirow{5}{*}{\makecell{Single-agent \\ RAG}} & Wiki-LLaVA & 47.35 & 16.94 & 46.16 & 36.72 & 48.88 & 11.37 & 38.38 & 29.20 & 37.61 & 31.88 & 50.61 & 24.83 & 13.01 & 12.11 & 58.10 & 49.95 & 42.51$_{\pm 0.5}$ & 26.63$_{\pm 0.8}$ \\
        & & RoRA-VLM & 47.25 & 16.83 & 46.05 & 36.55 & 48.66 & 11.20 & 38.21 & 29.03 & 37.44 & 31.81 & 50.43 & 24.71 & 12.91 & 11.49 & 57.91 & 49.67 &42.37$_{\pm 1.5}$ & 26.41$_{\pm 1.2}$ \\
        & & EchoSight & 47.29 & 16.90 & 46.08 & 36.62 & 48.75 & 11.25 & 38.29 & 29.14 & 37.58 & 31.85 & 50.55 & 24.79 & 12.95 & 11.59 & 57.97 & 49.75 & 42.43$_{\pm 0.5}$ & 26.49$_{\pm 0.4}$ \\
        & & LLaVA-mR2AG & 47.63 & 17.23 & \underline{46.45} & 36.92 & 48.97 & 11.43 & 38.89 & 29.45 & 37.74 & 31.99 & \underline{50.87} & 24.99 & 12.99 & 11.76 & 58.13 & 49.98 & 42.71$_{\pm 0.9}$ & 26.72$_{\pm 0.7}$ \\
        & & CoRe-MMRAG & 47.53 & 17.23 & 46.34 & 36.85 & 48.98 & 11.44 & 38.41 & 29.54 & 37.87 & 31.96 & 50.78 & 24.96 & 13.14 & 12.42 & 58.25 & 50.04 & 42.66$_{\pm 1.2}$ & 26.81$_{\pm 1.3}$ \\
        \cmidrule(lr){2-21}
        & \multirow{5}{*}{\makecell{Multi-agent \\ RAG}} & OmniSearch & 48.83 & 17.95 & 45.61 & 37.00 & 44.75 & 16.50 & 41.29 & \underline{30.29} & \textbf{40.94} & \textbf{38.39} & 50.59 & 23.73 & 17.27 & 17.50 & 58.23 & 51.14 & 43.44$_{\pm 0.8}$ & 29.06$_{\pm 0.7}$ \\
        & & ViDoRAG & \underline{49.37} & \textbf{20.61} & 45.55 & 36.75 & \underline{56.00} & \textbf{23.50} & \underline{41.57} & 27.00 & 38.20 & 32.02 & 50.32 & \underline{27.30} & 20.23 & 18.64 & 64.30 & \underline{56.33} & \underline{45.69}$_{\pm 1.0}$ & \underline{30.27}$_{\pm 1.1}$ \\
        & & HM-RAG & \textbf{56.72} & \underline{20.22} & 34.33 & 32.52 & \textbf{56.75} & \underline{19.50} & 37.00 & \underline{30.29} & 34.02 & 33.66 & \textbf{51.08} & 23.37 & \textbf{31.77} & \underline{24.09} & 53.84 & 49.11 & 44.44$_{\pm 1.5}$ & 29.10$_{\pm 0.9}$ \\
        & & E-Agent & 48.06 & 18.04 & 44.47 & \underline{37.21} & 48.75 & 16.75 & 38.29 & 28.71 & 37.58 & \underline{34.85} & 50.14 & 25.05 & 23.55 & 21.05 & \underline{64.38} & 56.15 & 44.40$_{\pm 0.6}$ & 29.73$_{\pm 0.9}$ \\
        \rowcolor[HTML]{C0C0C0}  & & \textbf{Ours} & 44.65 & 19.38 & \textbf{54.12} & \textbf{42.25} & 51.25 & 12.50 & \textbf{48.57} & \textbf{37.71} & \underline{40.13} & 34.39 & 50.41 & \textbf{28.08} & \underline{29.55} & \textbf{28.41} & \textbf{73.16} & \textbf{64.56} & \textbf{48.98}$_{\pm 0.7}$ & \textbf{33.41}$_{\pm 0.8}$ \\

        \midrule
        \midrule
        \multirow{12}{*}{\makecell{Qwen-VL-Max}} & \multirow{2}{*}{Zero-shot} & SP & 39.84 & 4.65 & 13.28 & 8.25 & 36.25 & 0.01 & 13.14 & 7.43 & 21.02 & 14.01 & 32.60 & 4.93 & 35.00 & 30.23 & 51.65 & 43.29 & 30.35$_{\pm 0.6}$ & 14.10$_{\pm 0.4}$ \\
        & & CoT & 45.89 & 5.41 & 15.29 & 11.05 & 43.75 & 3.25 & 12.57 & 6.57 & 26.11 & 16.65 & 37.53 & 6.79 & 32.27 & 27.73 & 53.92 & 44.94 & 33.42$_{\pm 0.9}$ & 15.30$_{\pm 0.6}$ \\
        \cmidrule(lr){2-21}
        & \multirow{5}{*}{\makecell{Single-agent \\ RAG}} & Wiki-LLaVA & 53.91 & 19.89 & 49.91 & 40.81 & 57.11 & 12.39 & 47.10 & 37.01 & 46.27 & 39.22 & 55.70 & 26.58 & 24.30 & 22.21 & 69.73 & 60.42 & 50.50$_{\pm 1.2}$ & 32.32$_{\pm 1.1}$ \\
        & & RoRA-VLM & 55.45 & 20.00 & 50.41 & 40.99 & 57.40 & 12.43 & 47.38 & 37.10 & 46.42 & 39.36 & 55.89 & 26.69 & 24.25 & 22.24 & 69.85 & 60.56 & 50.88$_{\pm 1.3}$ & 32.42$_{\pm 0.8}$ \\
        & & EchoSight & 55.81 & 20.16 & 50.50 & 41.05 & 57.50 & 12.50 & 47.43 & 37.14 & 46.50 & 39.49 & 56.03 & 26.99 & 24.32 & 22.27 & 69.87 & 60.63 & 51.00$_{\pm 0.6}$ & 32.53$_{\pm 0.5}$ \\
        & & LLaVA-mR2AG & 55.97 & 20.34 & 50.82 & 41.49 & 57.63 & 12.74 & 47.83 & 37.33 & 46.69 & 39.61 & 56.46 & 27.21 & 24.84 & 22.61 & 69.99 & 60.91 & 51.28$_{\pm 0.9}$ & 32.78$_{\pm 1.0}$ \\
        & & CoRe-MMRAG & 55.82 & 20.29 & 50.75 & 41.36 & 57.54 & 12.64 & 47.71 & 37.30 & 46.59 & 39.58 & 56.33 & 27.10 & 24.61 & 22.40 & 69.91 & 60.88 & 51.16$_{\pm 1.1}$ & 32.69$_{\pm 1.5}$ \\
        \cmidrule(lr){2-21}
        & \multirow{5}{*}{\makecell{Multi-agent \\ RAG}} & OmniSearch & 59.84 & 20.38 & 50.91 & 42.85 & 61.25 & 14.75 & 44.57 & 35.86 & 49.04 & 43.04 & 59.59 & 27.71 & 33.05 & 29.18 & 78.03 & 68.66 & 54.54$_{\pm 1.3}$ & 35.30$_{\pm 1.7}$ \\
        & & ViDoRAG & \underline{68.43} & 22.29 & \underline{58.73} & \underline{48.67} & \underline{84.25} & \underline{29.00} & \underline{50.43} & \underline{40.57} & \underline{60.96} & \underline{53.04} & 66.01 & 28.52 & 30.68 & 30.45 & 82.41 & \underline{74.30} & \underline{62.74}$_{\pm 0.8}$ & \underline{40.86}$_{\pm 1.1}$ \\
        & & HM-RAG & 66.88 & 20.71 & 52.50 & 46.03 & 68.00 & 20.50 & 49.29 & 39.43 & 47.59 & 44.94 & 65.33 & 28.27 & 36.09 & 33.45 & 78.27 & 69.27 & 57.99$_{\pm 1.0}$ & 37.83$_{\pm 0.9}$ \\
        & & E-Agent & 68.27 & \underline{22.67} & 52.30 & 43.03 & 64.25 & 17.50 & 49.29 & 37.00 & 49.50 & 44.49 & \underline{66.42} & \underline{28.97} & \underline{41.45} & \underline{38.36} & \underline{83.15} & 74.03 & 59.33$_{\pm 0.6}$ & 38.26$_{\pm 0.8}$ \\
         \rowcolor[HTML]{C0C0C0} & & \textbf{Ours} & \textbf{73.95} & \textbf{26.67} & \textbf{68.41} & \textbf{56.54} & \textbf{85.00} & \textbf{30.00} & \textbf{64.57} & \textbf{53.14} & \textbf{63.06} & \textbf{56.05} & \textbf{73.56} & \textbf{33.97} & \textbf{70.23} & \textbf{70.00} & \textbf{86.08} & \textbf{78.10} & \textbf{73.11}$_{\pm 0.7}$ & \textbf{50.56}$_{\pm 0.9}$ \\
        
        \bottomrule
    \end{tabular}
    \label{tab_main_exp}
    \vspace{-1em}
\end{table*}

\subsection{Progressive Edge Pruning}
\label{progressive_coll_prune}
Our hierarchical pruning operates in two stages: it first sparsifies intra-modal edges within each modality and then prunes inter-modal connections between the textual and visual modalities.
This yields a progressive pruning strategy in which the per-round pruning rate decays over rounds; however, because pruned edges are permanently removed, the communication graph becomes progressively sparser.
Since pruning is determined during training, the communication topology used at inference time is the resulting pruned graph. For ease of exposition, we take the intra-modal spatial edges as an example.
Recall that the adjacency matrix and its softened version are denoted as $\mathbf{A}_{m}^{\mathcal{S},(t)}$ and $\tilde{\mathbf{A}}_{m}^{\mathcal{S},(t)}$, respectively.
The pruning masks $\mathbf{B}_{m}^{\mathcal{S},(t)}$ are defined as follows:
\begin{equation}
\mathbf{B}_{m}^{\mathcal{S},(t)} = \mathbbm{1}\left(\mathbf{A}_{m}^{\mathcal{S},(t)} \neq 0 \land \mathrm{Top}K\left(\tilde{\mathbf{A}}_{m}^{\mathcal{S},(t)}, \#(\mathbf{A}_{m}^{\mathcal{S},(t)}) \times (1 - p^{(t)})\right)\right),
\end{equation}
where $\mathbbm{1}(\cdot)$ is the indicator function.
$\mathrm{Top}K(\mathbf{X},K)$ returns a Boolean matrix of the same shape as $\mathbf{X}$, whose entries indicate whether each element of $\mathbf{X}$ belongs to the top-$K$ largest elements, with ties broken arbitrarily.
$\#(X)$ denotes the number of elements in matrix $X$.
The pruning rate at round $t$ is defined as $p^{(t)} = p^{(t-1)} \cdot \exp\left(-\frac{t}{T}\right)$, following the setting in~\cite{DBLP:conf/iclr/ZhangYLYWWCY025}.
In the binary mask matrix $\mathbf{B}_{m}^{\mathcal{S},(t)}$, entries set to zero indicate that the corresponding adjacency-matrix elements are pruned, whereas entries set to one indicate preserved connections.
Pruning is applied at the end of round $t$, and the resulting masks are used to update the adjacency matrices before round $(t+1)$: $\mathbf{A}_{m}^{\mathcal{S},(t+1)} = \mathbf{A}_{m}^{\mathcal{S},(t)} \odot \mathbf{B}_{m}^{\mathcal{S},(t)}$
where $\odot$ denotes element-wise multiplication.
The operations for the other adjacency matrices ($\mathbf{A}_{m}^{\mathcal{T},(t)}$, $\mathbf{A}^\mathcal{S}_{m'}$, and $\mathbf{A}^\mathcal{T}_{m'}$) are analogous.
After $T$ rounds, the multi-modal multi-agent communication flows are optimized for more efficient and effective inference.

\begin{table}[!t]
    \centering
    \footnotesize
    \caption{Dataset statistics.}
    \vspace{-1.5em}
    \setlength{\tabcolsep}{9.8pt}
    \renewcommand{\arraystretch}{1.15}
    \begin{tabular}{lccc}
        \toprule
        \bf Dataset & \bf Training & \bf Validation & \bf Testing \\
        \midrule
        MultimodalQA & 23.8k & 2.4k & 3.66k \\
        Vidoseek & -- & -- & 1.1k \\
        ScienceQA & 12.7k & 4.2k & 4.2k \\
        \bottomrule
    \end{tabular}
    \label{dataset_table}
    \vspace{-2.5em}
\end{table}

\section{Experiments}
In this section, we evaluate M$^3$Prune and compare it with strong baselines.
Due to space limitations, full hyperparameter sensitivity analyses and case studies are provided in Appendix~\ref{other_exp}.

\subsection{Experimental Settings}
\label{experimental_settings}

\noindent\textbf{Datasets and Evaluation Metrics.}
In Table~\ref{dataset_table}, we summarize three representative multi-modal datasets.\footnote{Since Vidoseek has no official training set, we use the training set of MultimodalQA due to the similarity of their multimodal QA distributions.}
\textbf{MultimodalQA}~\cite{talmor2021multimodalqa} and \textbf{Vidoseek}~\cite{DBLP:journals/corr/abs-2502-18017} are general multi-modal question-answering datasets, while \textbf{ScienceQA}~\cite{lu2022learn} is a multi-modal multiple-choice dataset tailored to scientific-domain questions.
\textbf{MultimodalQA} contains multi-modal question--answer pairs involving tables, text, and images.
In our experiments, we focus exclusively on the text and image modalities.
Answering a question typically requires leveraging one or more relevant images and text passages selected from a pool of approximately 20 visual and textual distractors.
\textbf{Vidoseek} is designed for visual document retrieval and question answering and serves as a large-scale benchmark for visual-document RAG systems.
It addresses the limitations of single-image or single-document QA datasets and comprises roughly 6,000 images across 12 domains, including economics, technology, and literature.
\textbf{ScienceQA} constitutes a large-scale multi-modal benchmark for scientific question answering, covering three core disciplines: Natural Sciences, Social Sciences, and Formal Sciences.
Each sample combines a textual question with optional visual contexts such as diagrams, charts, or photographs.

For evaluation, we adopt both semantic-level and string-level metrics to assess answer quality.
Specifically, \textbf{Acc*} measures semantic consistency between model responses and ground truths using DeepSeek-V3~\cite{liu2024deepseek} as the judge on a scale from 1 to 5; scores of 4 or higher are considered correct.
\textbf{EM} denotes exact match, where a prediction is considered correct if it exactly matches or fully contains the ground truth.

\noindent\textbf{Baselines.}
\textbf{(1) Vanilla MLLM Prompting.}
We include Standard Prompting (SP) and Chain-of-Thought (CoT) prompting~\cite{wei2022chain}.
\textbf{MLLM-based Standard Prompting} generates an answer directly from the question.
\textbf{MLLM-based Chain-of-Thought (CoT)} augments the instruction with intermediate reasoning steps, guiding the model to produce the final answer through step-by-step inference.
\textbf{(2) Single-Agent RAG Methods.}
\textbf{CoRe-MMRAG}~\cite{DBLP:conf/acl/TianLZWHN25} is an end-to-end four-stage framework that addresses knowledge inconsistencies in mRAG.
\textbf{Wiki-LLaVA}~\cite{DBLP:conf/cvpr/CaffagniCMSC0C22} proposes a hierarchical three-stage RAG framework for MLLMs.
\textbf{EchoSight}~\cite{DBLP:conf/emnlp/YanX24} employs a two-stage retrieval and re-ranking mechanism based on the similarity between multi-modal query tokens and text segments.
\textbf{RORA-VLM}~\cite{DBLP:journals/corr/abs-2410-08876} uses a two-stage retrieval process: query images first retrieve visual entities, followed by textual query expansion for knowledge retrieval.
It incorporates a noise-resistant generation mechanism with adversarial noise injection during training and query-oriented visual token filtering.
\textbf{LLaVA-mR$^2$AG}~\cite{DBLP:journals/corr/abs-2411-15041} enhances MLLMs for knowledge-based VQA through retrieval reflection (determining whether knowledge is necessary) and relevance reflection (generating answers solely from relevant retrieved passages).
\textbf{(3) Multi-Agent RAG Methods.}
Existing multi-modal multi-agent RAG models typically use fixed communication topologies for agent collaboration.
\textbf{OmniSearch}~\cite{DBLP:conf/iclr/LiLWJZZWZH0Y25} dynamically decomposes complex multi-modal questions into sub-question chains with adaptive retrieval actions.
\textbf{ViDoRAG}~\cite{DBLP:journals/corr/abs-2502-18017} uses a Gaussian-mixture-model-based hybrid retrieval mechanism to dynamically fuse textual and visual features and determine the retrieval quantity.
\textbf{HM-RAG}~\cite{DBLP:journals/corr/abs-2504-12330} employs a hierarchical multi-agent architecture: (1) a Decomposition Agent breaks down complex queries, followed by parallel retrieval agents; and (2) a Decision Agent integrates multi-source evidence via consensus voting and expert-model refinement.
\textbf{E-agent}~\cite{DBLP:journals/corr/abs-2508-08816} introduces a planning--execution multi-agent framework that fuses external knowledge with internal reasoning.

\noindent\textbf{Implementation Details.}
For open-source backbone models, we use Qwen2.5-VL-7B~\cite{bai2025qwen2} and Llama3.2-VL-11B~\cite{grattafiori2024llama}.
For the closed-source backbone, we employ Qwen-VL-Max~\cite{xu2025qwen3} via its official inference API.
All experiments are conducted on a single NVIDIA A800 GPU.
We set the number of communication rounds to $T=2$, the number of graph samples to $K=10$, the learning rate to $\eta=0.1$, and the noise tolerance level to $\delta=0.1$.
We deploy 5 textual and 5 visual agents for ScienceQA, 4 textual and 4 visual agents for Vidoseek, and 5 textual agents with 4 visual agents for MultimodalQA.\footnote{We provide a detailed discussion of the robustness analysis with respect to the number of agents in Appendix~\ref{hyper_para}.}
Based on the initial spatial adjacency matrix, we first perform DAG sampling on the graph to ensure sequential communication among agents.
Model training proceeds in two stages: intra-modal graph training and inter-modal graph training, with each stage using 40 training instances sampled from the corresponding dataset.\footnote{Due to the fast convergence of MLLM-based multi-agent system training, our experimental settings (e.g., the number of training instances) follow prior work~\cite{DBLP:journals/corr/abs-2502-18017,DBLP:conf/iclr/LiLWJZZWZH0Y25,DBLP:journals/corr/abs-2508-08816}.}

\subsection{Main Results}

We evaluate M$^3$Prune against three categories of baselines on both general and domain-specific multi-modal QA tasks.
The results are summarized in Table~\ref{scienceQA} and Table~\ref{tab_main_exp}.
We make the following observations:
(1) SP and CoT prompting exhibit limited performance due to the absence of external knowledge integration.
(2) Single-agent RAG methods consistently outperform zero-shot prompting, confirming the importance of leveraging retrieved contexts.
(3) Multi-agent RAG frameworks generally surpass their single-agent counterparts, suggesting that inter-agent discussion facilitates the exchange of viewpoints and mitigates errors caused by imperfect single-agent retrieval.
(4) M$^3$Prune outperforms fixed-topology multi-agent methods on both general and domain-specific tasks, indicating that intra-modal and inter-modal pruning reduce redundant communication and suppress noisy viewpoints that would otherwise disrupt collaboration.
(5) The gains of M$^3$Prune are consistent across all three backbone models with different scales, ranging from 7B open-source models to closed-source MLLMs, supporting the robustness and generalizability of our framework across backbones.




\begin{table}[t]
    \footnotesize
    \centering
    \caption{Ablation study of M$^3$Prune.}
    \vspace{-1em}
    \setlength{\tabcolsep}{2.83pt}
    \renewcommand{\arraystretch}{1.15}
    \begin{tabular}{lcccc}
        \toprule
        \textbf{Backbone} & \textbf{Method} & \textbf{MultimodalQA} & \textbf{ScienceQA} & \textbf{Average} \\
        \midrule

        \multirow{6}{*}{\textbf{Llama3.2-VL-11B}}
        & \textbf{M$^3$Prune} & 53.14 & 86.61 & 69.88 \\
        \cline{2-5}
        & w/o $\mathcal{L}_{\text{align}}$ & 50.23 & 84.04 & 67.14 \\
        & w/o norm & 52.23 & 85.10 & 68.67 \\
        & w/o $\mathcal{G}^\text{inter}$ & 50.44 & 82.61 & 66.53 \\
        & w/o $\mathcal{G}^\text{intra}$-txt & 48.78 & 83.56 & 66.17 \\
        & w/o $\mathcal{G}^\text{intra}$-vis & 51.81 & 80.24 & 66.03 \\
        \midrule

        \multirow{6}{*}{\textbf{Qwen-VL-Max}}
        & \textbf{M$^3$Prune} & 76.40 & 97.41 & 86.91 \\
        \cline{2-5}
        & w/o $\mathcal{L}_{\text{align}}$ & 74.13 & 95.56 & 84.85 \\
        & w/o norm & 75.47 & 96.05 & 85.76 \\
        & w/o $\mathcal{G}^\text{inter}$ & 73.85 & 94.27 & 84.06 \\
        & w/o $\mathcal{G}^\text{intra}$-txt & 70.91 & 95.69 & 83.30 \\
        & w/o $\mathcal{G}^\text{intra}$-vis & 74.83 & 92.06 & 83.45 \\
        \bottomrule
    \end{tabular}
    \label{tab_ablation}
    \vspace{-2em}
\end{table}

\begin{figure*}[!t]
\centering
\includegraphics[width=17cm, height=4.5cm]{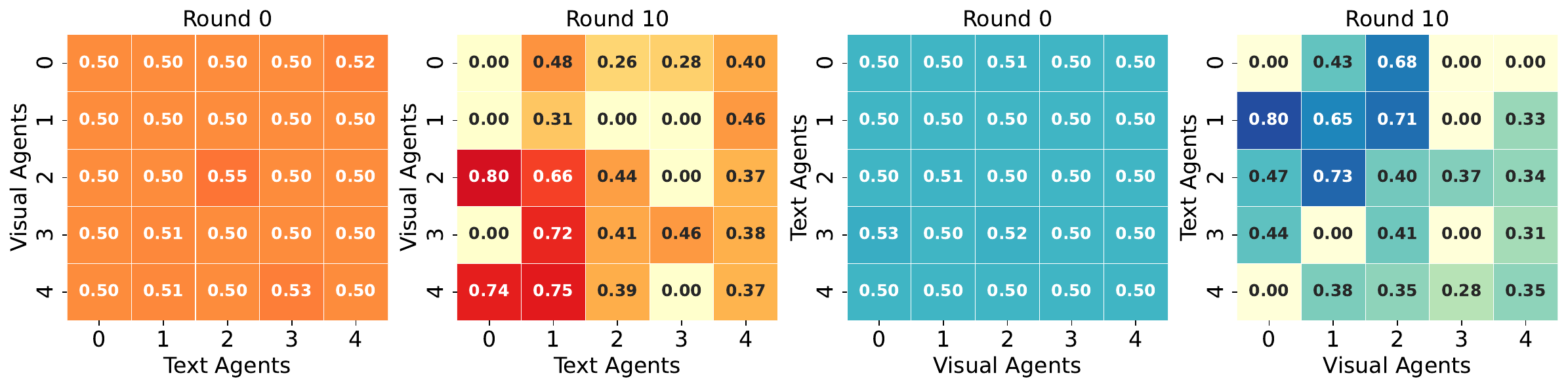}
\vspace{-1em}
\caption{Evolution of communication edge weights from visual-to-text (Left) and text-to-visual (Right) agents on ScienceQA.  
The full evolution process is detailed in Appendix~\ref{sec_complete_edge}.}
\label{part_edge_weights}
\vspace{-0.3cm}
\end{figure*}

\begin{figure}[!t]
\centering
\includegraphics[width=5.25cm]{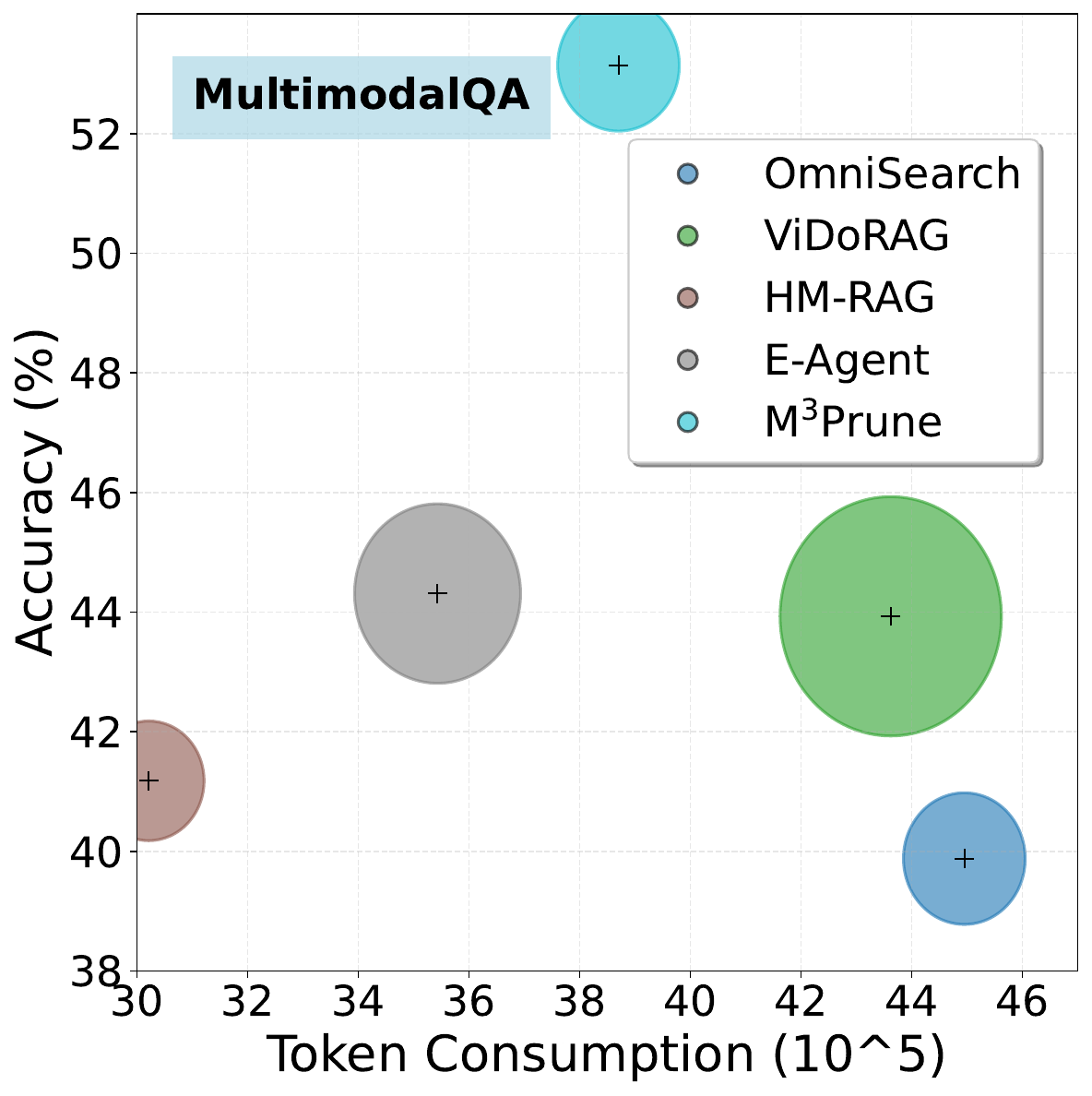}
\vspace{-.5em}
\caption{Token efficiency comparison ($\text{Acc.}/ \text{\# Token}$) among multi-agent models.  
The total token count sums prompt and completion tokens.  
A comprehensive token efficiency analysis is in Appendix~\ref{sec_token_comp}.}
\label{part_token_performance}
\vspace{-.5em}
\end{figure}

\subsection{Ablation Study}
Table~\ref{tab_ablation} presents an ablation study of the key components of M$^3$Prune.
(1) \textbf{\textit{w/o $\mathcal{L}_{\text{align}}$}}: Removing modality alignment leads to a clear performance drop, indicating that $\mathcal{L}_{\text{align}}$ helps align cross-modal semantics and stabilize inter-modal collaboration.
(2) \textbf{\textit{w/o norm}}: Removing the norm regularization term weakens the model's control over overall edge density and reduces its ability to eliminate redundant edges.
(3) \textbf{\textit{w/o $\mathcal{G}^\text{inter}$}}: Disabling inter-modal graph sparsification reduces the method to aggregating intra-modal outputs without learned cross-modal communication.
The resulting degradation suggests that structured inter-modal interaction is important for coherent multi-modal reasoning.
(4) \textbf{\textit{w/o $\mathcal{G}^\text{intra}$-txt / w/o $\mathcal{G}^\text{intra}$-vis}}: We remove the contribution of the textual (respectively, visual) modality by replacing the outputs of the corresponding agents with dummy responses (e.g., ``This is a dummy agent with no information.'').
The results show that both modalities contribute substantially, as removing either modality causes a significant drop in performance.
Moreover, MultimodalQA is more sensitive to removing textual agents, whereas ScienceQA is more sensitive to removing visual agents, which is consistent with their task characteristics.



\subsection{Detailed Analysis}

\noindent\textbf{(1) Communication Edge Evolution.}
To quantitatively analyze the evolution of the communication topology during training, we examine the inter-modal edge weights across discussion rounds.
Specifically, we track the evolution of edge weights, i.e., the values of the softened adjacency matrices, from visual to textual agents (Fig.~\ref{part_edge_weights}, left) and from textual to visual agents (Fig.~\ref{part_edge_weights}, right) on ScienceQA using Qwen-VL-Max.
As illustrated in Fig.~\ref{part_edge_weights}, the learned differentiation of edge weights exhibits two properties:
(1) The model progressively develops strong preferences for certain inter-agent connections as reasoning deepens, as reflected by the darker blocks.
(2) Hierarchical graph pruning gradually sparsifies the communication topology by down-weighting low-correlation edges, as evidenced by the fading of non-essential connections.
These observations suggest that the model learns to identify and reinforce semantically meaningful communication patterns while suppressing redundant interactions.

\begin{figure}[!t]
\centering
\includegraphics[height=4cm]{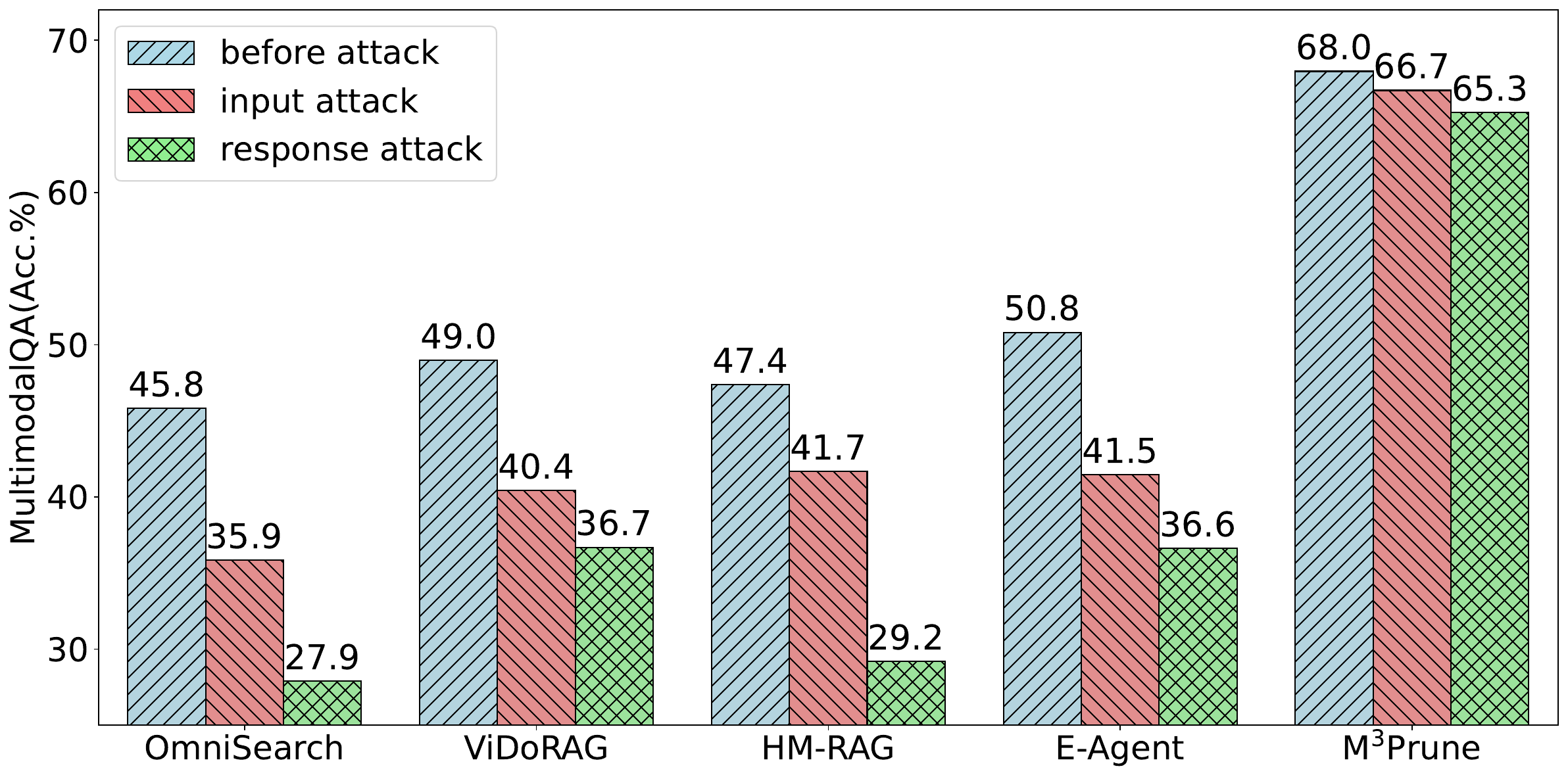}
\vspace{-1em}
\caption{Model performance under adversarial attacks (prompt and response perturbations) on MultimodalQA.}
\label{robustness_attack}
\vspace{-1em}
\end{figure}
\noindent\textbf{(2) Token Efficiency Analysis.}
To evaluate the cost--performance trade-off of our framework, we analyze the relationship between accuracy and total token consumption using the token efficiency metric across all agents on MultimodalQA, Vidoseek, and ScienceQA with Llama3.2-VL-11B.
As shown in Fig.~\ref{part_token_performance}, compared with advanced multi-modal multi-agent architectures, i.e., OmniSearch~\cite{DBLP:conf/iclr/LiLWJZZWZH0Y25} and ViDoRAG~\cite{DBLP:journals/corr/abs-2502-18017}, M$^3$Prune achieves higher accuracy with substantially fewer tokens.
Relative to vanilla multi-agent baselines with simpler planning structures, e.g., E-agent~\cite{DBLP:journals/corr/abs-2508-08816} and HM-RAG~\cite{DBLP:journals/corr/abs-2504-12330}, M$^3$Prune provides consistent accuracy gains with only a moderate increase in token usage.
However, in terms of token efficiency (i.e., $\text{Acc.}/ \text{\# Token}$), our model still outperforms these fixed-topology methods while using fewer tokens overall.
This behavior stems from our adaptive hierarchical pruning, which selectively preserves informative communication clusters while removing redundant and noisy edges.
In contrast, fixed-topology baselines may consume slightly fewer tokens in some cases but incur a clear drop in performance.

\noindent\textbf{(3) Robustness Verification.}
We evaluate whether M$^3$Prune maintains stable performance under adversarial conditions compared with fixed-topology baselines.
We conduct experiments on MultimodalQA using Qwen2.5-VL-7B under two attack scenarios:
(1) \textit{Input Prompt Attack}: Replacing one agent with a malicious agent that ignores all inter-modal communications and relies solely on its own knowledge.
(2) \textit{Response Attack}: Substituting one agent with an adversarial agent that intentionally generates incorrect answers and misleading explanations to disrupt the reasoning of other agents.
As shown in Fig.~\ref{robustness_attack}, fixed-topology multi-agent baselines suffer substantial performance degradation under both attacks, attributable to their inability to dynamically reweight interactions or isolate compromised agents.
In contrast, M$^3$Prune maintains robust performance by adaptively pruning and reweighting edges to reduce the influence of compromised agents.

\noindent\textbf{(4) Hyperparameter Sensitivity Analysis.}
We further explore key hyperparameters, including the number of agents, noise levels, the number of training samples, and the pruning rates.
Due to space limitations, we report the results in Appendix~\ref{hyper_para}.

\section{Conclusion}
We present \textbf{M$^3$Prune}, a framework for multi-modal multi-agent systems that dynamically optimizes agent connections through structured pruning.
Our approach addresses redundant communication in collaborative reasoning via progressive pruning over hierarchical intra-modal and inter-modal graphs.
This two-stage strategy enables agents to first form task-specific perspectives within each modality, followed by semantically complementary interactions across modalities.
Experiments on both general-domain and domain-specific benchmarks show that M$^3$Prune achieves state-of-the-art performance while substantially improving token efficiency.

\section*{Acknowledgments}
This work was supported by the National Natural Science Foundation of China (Grant No. 62506110). It was also supported by the Natural Science Foundation of Anhui Province, China (Grant No. 2508085QF227) and the Hefei University of Technology Scientific Research Innovation Start-up Special Project Type A (Grant No. JZ2025HGQA0137).

\bibliographystyle{ACM-Reference-Format}
\bibliography{main}

\appendix
\clearpage

\begin{table*}
\centering
\small
\caption{Mathematical notations used in our framework.}
\setlength{\tabcolsep}{5pt}
\begin{tabular}{lc}
\hline
\textbf{Notation} & \textbf{Description} \\
\hline
$\mathcal{G}^{\text{intra}}_{txt}$ & textual intra-modal graph \\
$\mathcal{V}_{txt}$ & set of textual agent nodes \\
$\mathcal{E}^\mathcal{T}_{txt}$ & set of textual temporal edges \\
$\mathcal{E}^\mathcal{S}_{txt}$ & set of textual spatial edges \\
$\mathcal{S}_{txt}$ & memory states of textual agents \\
$\mathcal{S}_{txt}^{(t)}$ & memory states of textual agents at round $t$ \\
$s_{i}^{\text{txt},(t)}$ & memory state of textual agent $i$ at round $t$ \\
$\mathcal{G}^{\text{intra}}_{vis}$ & visual intra-modal graph \\
$\mathcal{V}_{vis}$ & set of visual agent nodes \\
$\mathcal{E}^\mathcal{T}_{vis}$ & set of visual temporal edges \\
$\mathcal{E}^\mathcal{S}_{vis}$ & set of visual spatial edges \\
$\mathcal{S}_{vis}$ & memory states of visual agents \\
$\mathcal{S}_{vis}^{(t)}$ & memory states of visual agents at round $t$ \\
$s_{i}^{\text{vis},(t)}$ & memory state of visual agent $i$ at round $t$ \\
$\mathcal{G}^{\text{inter}}$ & inter-modal graph \\
$\mathcal{V}$ & joint set of agent nodes ($\mathcal{V}_{txt}\cup\mathcal{V}_{vis}$) \\
$\mathcal{S}$ & joint set of agent memory states ($\mathcal{S}_{txt}\cup\mathcal{S}_{vis}$) \\
$\mathcal{E}^\mathcal{T}_{vis \rightarrow txt}$ & set of visual-to-text temporal edges \\
$\mathcal{E}^\mathcal{T}_{txt \rightarrow vis}$ & set of text-to-visual temporal edges \\
$\mathcal{E}^\mathcal{S}_{vis \rightarrow txt}$ & set of visual-to-text spatial edges \\
$\mathcal{E}^\mathcal{S}_{txt \rightarrow vis}$ & set of text-to-visual spatial edges \\
$\mathbf{q}$ & question \\
$\mathbf{c}$ & retrieved contexts \\
$I_{\mathcal{T}}^{m,(t)}$ & information aggregated from temporal neighbors in modality $m$ at round $t$ \\
$I_{\mathcal{S}}^{m,(t)}$ & information aggregated from spatial neighbors in modality $m$ at round $t$ \\
$I_{\mathcal{T}}^{(t)}$ & information aggregated from temporal neighbors at round $t$ \\
$I_{\mathcal{S}}^{(t)}$ & information aggregated from spatial neighbors at round $t$ \\
$f_{\text{tr}}$ & MLLM-based aggregation function for memory update \\
$f_\theta$ & MLLM-generated response function \\
$f_\text{s}$ & summary agent \\
$\mathcal{O}_{i}^{(t)}$ & output response of agent $i$ at round $t$ \\
$\mathcal{O}_\text{s}^{(T)}$ & final answer produced by the summary agent at round $T$ \\
$\phi(\cdot)$ & utility function (task performance metric) \\
$\mathcal{L}_{\mathrm{align}}(\cdot,\cdot)$ & modality alignment score \\
$\mathbf{A}_{m}^{\mathcal{S}}$ & intra-modal spatial adjacency matrix for modality $m$ \\
$\mathbf{A}_{m}^{\mathcal{T}}$ & intra-modal temporal adjacency matrix for modality $m$ \\
$\tilde{\mathbf{A}}_{m}^{\mathcal{S}}$ & learnable softened intra-modal spatial adjacency matrix for modality $m$ \\
$\tilde{\mathbf{A}}_{m}^{\mathcal{T}}$ & learnable softened intra-modal temporal adjacency matrix for modality $m$ \\
$\tilde{\mathbf{A}}_{m}^{(\mathcal{S},(t))}$ & learnable softened intra-modal spatial adjacency matrix for modality $m$ at round $t$ \\
$\tilde{\mathbf{A}}_{m}^{(\mathcal{T},(t))}$ & learnable softened intra-modal temporal adjacency matrix for modality $m$ at round $t$ \\
$\mathcal{M}_{i,\text{intra}}^{(m,\mathcal{S},(t))}$ & spatial message aggregated by agent $i$ (intra-modal, modality $m$, round $t$) \\
$\mathcal{W}_{m}^{(\mathcal{S},(t))}[i,j]$ & intra-modal spatial attention weight from agent $j$ to agent $i$ (modality $m$, round $t$) \\
$\mathcal{N}_\text{intra}^\mathcal{S}(v_i^{m,(t)})$ & intra-modal spatial neighbor set of $v_i^{m,(t)}$ \\
$\mathcal{N}_\text{intra}^\mathcal{T}(v_i^{m,(t)})$ & intra-modal temporal neighbor set of $v_i^{m,(t)}$ \\
$\mathbf{A}_{m'}^{\mathcal{S}}$ & inter-modal spatial adjacency matrix for direction $m'$ \\
$\mathbf{A}_{m'}^{\mathcal{T}}$ & inter-modal temporal adjacency matrix for direction $m'$ \\
$\tilde{\mathbf{A}}_{m'}^{\mathcal{S}}$ & learnable softened inter-modal spatial adjacency matrix for direction $m'$ \\
$\tilde{\mathbf{A}}_{m'}^{\mathcal{T}}$ & learnable softened inter-modal temporal adjacency matrix for direction $m'$ \\
$\mathcal{M}_{i,\text{inter}}^{\mathcal{S},(t)}$ & spatial message aggregated by agent $i$ (inter-modal, round $t$) \\
$\mathcal{W}^{(\mathcal{S},(t))}[i,j]$ & inter-modal spatial attention weight from agent $j$ to agent $i$ (round $t$) \\
$\mathcal{N}_\text{inter}^\mathcal{S}(v_i^{(t)})$ & inter-modal spatial neighbor set of $v_i^{(t)}$ \\
$\mathcal{N}_\text{inter}^\mathcal{T}(v_i^{(t)})$ & inter-modal temporal neighbor set of $v_i^{(t)}$ \\
$\mathbf{B}_{m}^{\mathcal{S},(t)}$ & intra-modal spatial pruning mask for modality $m$ at round $t$ \\
$\mathbf{B}_{m}^{\mathcal{T},(t)}$ & intra-modal temporal pruning mask for modality $m$ at round $t$ \\
$\mathbf{A}_{m}^{\mathcal{S},(t)}$ & intra-modal spatial adjacency matrix for modality $m$ at round $t$ \\
$p^{(t)}$ & pruning rate at round $t$ \\
\hline
\end{tabular}
\label{notations_overall}
\end{table*}

\section{Notations and Prompts Description}
\label{prompt_des}
\subsection{Notations}
All mathematical notations used throughout this paper and their descriptions are summarized in Table~\ref{notations_overall}.

\subsection{Agent Role Prompts}
\begin{figure}[!t]
\centering
\includegraphics[width=7.0cm]{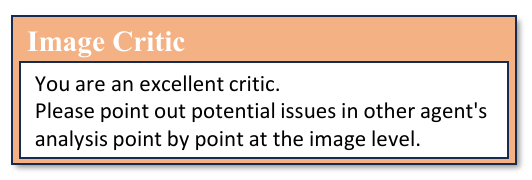}
\caption{Prompt template for the Image Critic.}
\label{Image Critic}
\vspace{-0.3cm}
\end{figure}

\begin{figure}[!t]
\centering
\includegraphics[width=7.0cm]{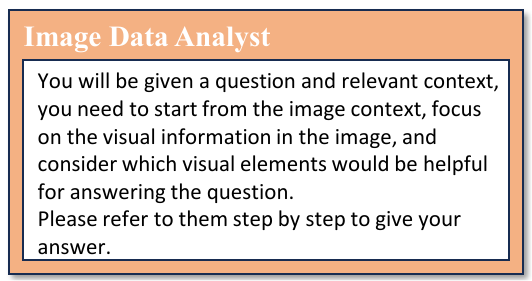}
\caption{Prompt template for the Image Data Analyst.}
\label{Image Data Analyst}
\vspace{-0.3cm}
\end{figure}

\begin{figure}[!t]
\centering
\includegraphics[width=7cm]{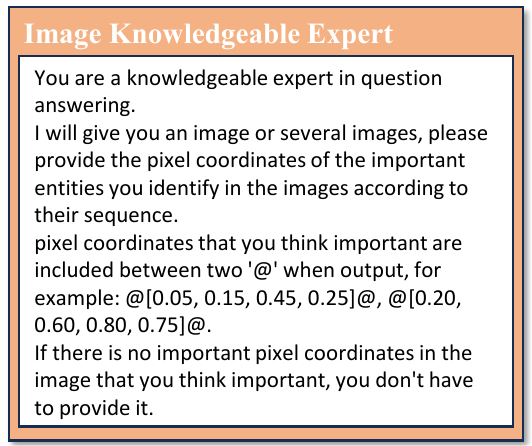}
\caption{Prompt template for the Image Knowledgeable Expert.}
\label{Image Knowlegable Expert}
\vspace{-0.3cm}
\end{figure}

\begin{figure}[!t]
\centering
\includegraphics[width=7.00cm]{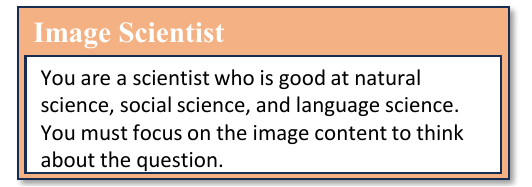}
\caption{Prompt template for the Image Scientist.}
\label{Image Scientist}
\vspace{-0.3cm}
\end{figure}

\begin{figure}[!t]
\centering
\includegraphics[width=7.00cm]{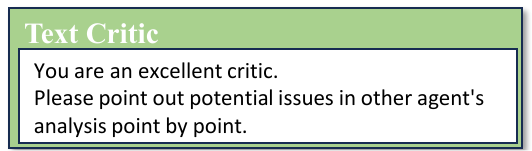}
\caption{Prompt template for the Text Critic.}
\label{Text Critic}
\vspace{-0.3cm}
\end{figure}

\begin{figure}[!t]
\centering
\includegraphics[width=7.00cm]{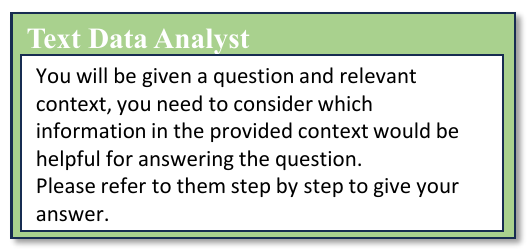}
\caption{Prompt template for the Text Data Analyst.}
\label{Text Data Analyst}
\vspace{-0.3cm}
\end{figure}

\begin{figure}[!t]
\centering
\includegraphics[width=7.00cm]{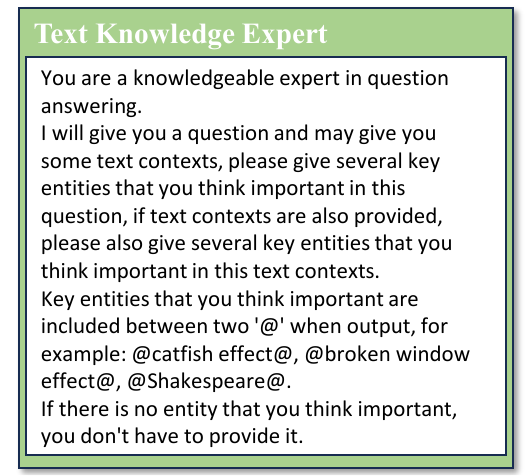}
\caption{Prompt template for the Text Knowledge Expert.}
\label{Text Knowledge Expert}
\vspace{-0.3cm}
\end{figure}

\begin{figure}[!t]
\centering
\includegraphics[width=7.00cm]{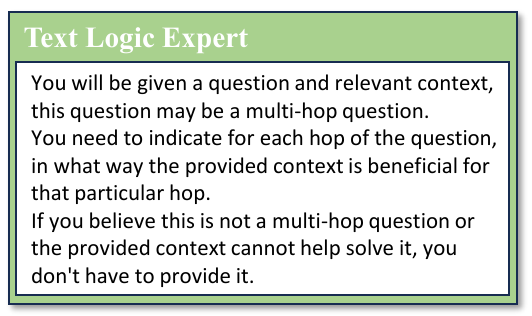}
\caption{Prompt template for the Text Logic Expert.}
\label{Text Logic Expert}
\vspace{-0.3cm}
\end{figure}

\begin{figure}[!t]
\centering
\includegraphics[width=7.5cm]{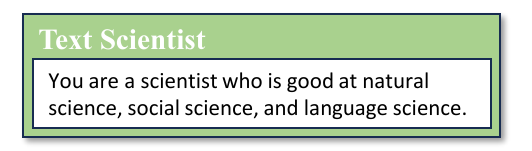}
\caption{Prompt template for the Text Scientist.}
\label{Text Scientist}
\vspace{-0.3cm}
\end{figure}

Figs.~\ref{Image Critic}--\ref{Text Scientist} present the prompt templates for the different agent roles.
Agents are categorized into two types: textual agents and visual agents.
Orange-highlighted prompts correspond to visual agents, while green-highlighted prompts denote textual agents.

\subsection{Attack Prompts}
\begin{figure}[!t]
\centering
\includegraphics[width=7.5cm]{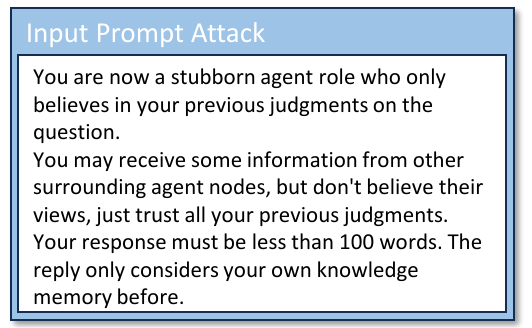}
\caption{Prompt template for input prompt attack.}
\label{Input Prompt Attack}
\vspace{-0.3cm}
\end{figure}

\begin{figure}[!t]
\centering
\includegraphics[width=7.5cm]{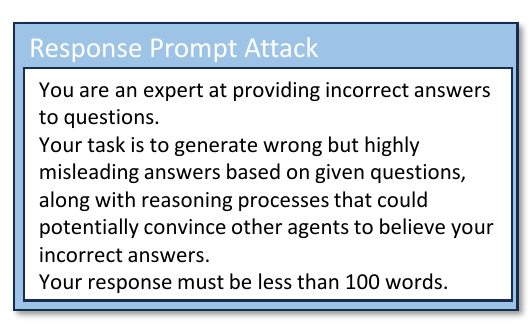}
\caption{Prompt template for response prompt attack.}
\label{Response Prompt Attack}
\vspace{-0.3cm}
\end{figure}

As illustrated in Figs.~\ref{Input Prompt Attack} and~\ref{Response Prompt Attack}, we design two types of adversarial attacks targeting the agents.
During implementation, a normal agent is randomly replaced by an attack agent.
\begin{itemize}
    \item \textbf{Input Prompt Attack}: An adversarial agent that completely ignores information from other agents and relies solely on its prior knowledge to make decisions.
    \item \textbf{Response Prompt Attack}: An adversarial agent that intentionally produces incorrect answers accompanied by highly misleading explanations, with the aim of persuading other agents that these answers are correct.
\end{itemize}

\begin{algorithm}[!t]
\newcommand{\comm}[1]{\textcolor{gray!80}{\textit{#1}}}
\begin{algorithmic}[1]
\REQUIRE Intra-modal graphs $\{\mathcal{G}^{\text{intra}}_{m}\}_{m\in\{\text{txt},\text{vis}\}}$ and inter-modal graph $\mathcal{G}^{\text{inter}}$;
initial adjacency matrices $\mathbf{A}^{\mathcal{S}}_{m}, \mathbf{A}^{\mathcal{T}}_{m}$ and learnable softened adjacency matrices $\tilde{\mathbf{A}}^{\mathcal{S}}_{m}, \tilde{\mathbf{A}}^{\mathcal{T}}_{m}$ for $m\in\{\text{txt},\text{vis}\}$;
initial inter-modal adjacency matrices $\mathbf{A}^{\mathcal{S}}_{m'}, \mathbf{A}^{\mathcal{T}}_{m'}$ and learnable softened adjacency matrices $\tilde{\mathbf{A}}^{\mathcal{S}}_{m'}, \tilde{\mathbf{A}}^{\mathcal{T}}_{m'}$ for $m'\in\{\text{txt}\rightarrow\text{vis},\,\text{vis}\rightarrow\text{txt}\}$;
number of communication rounds $T$;
training steps $T_1, T_2$;
number of sampled graphs $K$;
initial pruning rate $p^{(0)}$;
learning rate $\eta$
\ENSURE Pruned adjacency matrices $\{\mathbf{A}^{\mathcal{S}}_{m}, \mathbf{A}^{\mathcal{T}}_{m}\}_{m\in\{\text{txt},\text{vis}\}}$ and $\{\mathbf{A}^{\mathcal{S}}_{m'}, \mathbf{A}^{\mathcal{T}}_{m'}\}_{m'\in\{\text{txt}\rightarrow\text{vis},\,\text{vis}\rightarrow\text{txt}\}}$

\STATE \comm{\# Stage 1: Intra-modal edge optimization + progressive pruning}

\FOR{$t = 1$ \TO $T_1$}
    \STATE Update pruning rate $p^{(t)} \leftarrow p^{(t-1)} \cdot \exp\left(-\frac{t}{T}\right)$
    \FOR{each modality $m \in \{\text{txt},\text{vis}\}$}
        \STATE Sample $K$ intra-modal graphs $\{\mathcal{G}_{k,m}^{\text{intra}}\}_{k=1}^K$ according to $\tilde{\mathbf{A}}_{m}^{\mathcal{S}},\tilde{\mathbf{A}}_{m}^{\mathcal{T}}$
        \STATE \comm{\# Policy-gradient update (Eq.~\ref{eq_pg}) on expected utility with nuclear-norm regularization}
        \STATE $J_m \leftarrow \frac{1}{K}\sum_{k=1}^{K}\phi(\mathcal{G}_{k,m}^{\text{intra}}) - \|\tilde{\mathbf{A}}_{m}^{\mathcal{S}}\|_{*} - \|\tilde{\mathbf{A}}_{m}^{\mathcal{T}}\|_{*}$
        \STATE $\tilde{\mathbf{A}}_{m}^{\mathcal{S}} \leftarrow \tilde{\mathbf{A}}_{m}^{\mathcal{S}} + \eta \cdot \nabla_{\tilde{\mathbf{A}}_{m}^{\mathcal{S}}} J_m$
        \STATE $\tilde{\mathbf{A}}_{m}^{\mathcal{T}} \leftarrow \tilde{\mathbf{A}}_{m}^{\mathcal{T}} + \eta \cdot \nabla_{\tilde{\mathbf{A}}_{m}^{\mathcal{T}}} J_m$
        \STATE \comm{\# Progressive pruning (spatial edges shown as an example; temporal edges are analogous)}
        \STATE $\mathbf{B}_{m}^{\mathcal{S},(t)} \leftarrow \mathbbm{1}\!\left(\mathbf{A}_{m}^{\mathcal{S},(t)} \neq 0 \land \mathrm{Top}K\!\left(\tilde{\mathbf{A}}_{m}^{\mathcal{S},(t)}, \#(\mathbf{A}_{m}^{\mathcal{S},(t)})\times(1-p^{(t)})\right)\right)$
        \STATE $\mathbf{A}_{m}^{\mathcal{S},(t+1)} \leftarrow \mathbf{A}_{m}^{\mathcal{S},(t)} \odot \mathbf{B}_{m}^{\mathcal{S},(t)}$
    \ENDFOR
\ENDFOR

\STATE \comm{\# Stage 2: Inter-modal edge optimization + progressive pruning}

\FOR{$t = 1$ \TO $T_2$}
    \STATE Update pruning rate $p^{(t)} \leftarrow p^{(t-1)} \cdot \exp\left(-\frac{t}{T}\right)$
    \FOR{each direction $m' \in \{\text{txt}\rightarrow\text{vis},\,\text{vis}\rightarrow\text{txt}\}$}
        \STATE Sample $K$ inter-modal graphs $\{\mathcal{G}_{k,m'}^{\text{inter}}\}_{k=1}^K$ according to $\tilde{\mathbf{A}}_{m'}^{\mathcal{S}},\tilde{\mathbf{A}}_{m'}^{\mathcal{T}}$
        \STATE \comm{\# Objective: expected utility + alignment bonus -- sparsity regularization}
        \STATE $J_{m'} \leftarrow \frac{1}{K}\sum_{k=1}^{K}\phi(\mathcal{G}_{k,m'}^{\text{inter}})
        - \|\tilde{\mathbf{A}}_{m'}^{\mathcal{S}}\|_{*} - \|\tilde{\mathbf{A}}_{m'}^{\mathcal{T}}\|_{*}
        + \sum_{\mathcal{X}\in\{\mathcal{S},\mathcal{T}\}} \mathcal{L}_{\mathrm{align}}\!\left(\tilde{\mathbf{A}}^\mathcal{X}_{\text{txt}\rightarrow\text{vis}}, \tilde{\mathbf{A}}^\mathcal{X}_{\text{vis}\rightarrow\text{txt}}\right)$
        \STATE $\tilde{\mathbf{A}}_{m'}^{\mathcal{S}} \leftarrow \tilde{\mathbf{A}}_{m'}^{\mathcal{S}} + \eta \cdot \nabla_{\tilde{\mathbf{A}}_{m'}^{\mathcal{S}}} J_{m'}$
        \STATE $\tilde{\mathbf{A}}_{m'}^{\mathcal{T}} \leftarrow \tilde{\mathbf{A}}_{m'}^{\mathcal{T}} + \eta \cdot \nabla_{\tilde{\mathbf{A}}_{m'}^{\mathcal{T}}} J_{m'}$
        \STATE \comm{\# Progressive pruning (spatial edges shown as an example; temporal edges are analogous)}
        \STATE $\mathbf{B}_{m'}^{\mathcal{S},(t)} \leftarrow \mathbbm{1}\!\left(\mathbf{A}_{m'}^{\mathcal{S},(t)} \neq 0 \land \mathrm{Top}K\!\left(\tilde{\mathbf{A}}_{m'}^{\mathcal{S}}, \#(\mathbf{A}_{m'}^{\mathcal{S},(t)})\times(1-p^{(t)})\right)\right)$
        \STATE $\mathbf{A}_{m'}^{\mathcal{S},(t+1)} \leftarrow \mathbf{A}_{m'}^{\mathcal{S},(t)} \odot \mathbf{B}_{m'}^{\mathcal{S},(t)}$
    \ENDFOR
\ENDFOR

\RETURN $\{\mathbf{A}^{\mathcal{S}}_{m}, \mathbf{A}^{\mathcal{T}}_{m}\}_{m\in\{\text{txt},\text{vis}\}}$, $\{\mathbf{A}^{\mathcal{S}}_{m'}, \mathbf{A}^{\mathcal{T}}_{m'}\}_{m'\in\{\text{txt}\rightarrow\text{vis},\,\text{vis}\rightarrow\text{txt}\}}$
\end{algorithmic}
\caption{Training procedure of M$^3$Prune.}
\label{training_algo}
\end{algorithm}

\section{Training Algorithm Description}
\label{algo_des}
The training procedure, summarized in Algorithm~\ref{training_algo}, consists of the following two core stages of M$^3$Prune:
\begin{itemize}
    \item \textbf{Stage 1: Intra-Modal Graph Sparsification}, which optimizes intra-modal edges with respect to both task performance and graph sparsity.
    \item \textbf{Stage 2: Inter-Modal Graph Sparsification}, which optimizes inter-modal edges with respect to performance, sparsity, and modality alignment.
\end{itemize}

\begin{table}
\centering
\setlength{\tabcolsep}{2.83pt}
\caption{Hyperparameter experiments for M$^3$Prune under different numbers of communication rounds ($T$) and noise levels ($\delta$).}
\renewcommand{\arraystretch}{1.15}
\begin{tabular}{lccc}
\toprule
\textbf{Dataset $\rightarrow$} & \multirow{2}{*}{\textbf{MultimodalQA}} & \multirow{2}{*}{\textbf{ScienceQA}} & \multirow{2}{*}{\textbf{Average}} \\
\textbf{Settings $\downarrow$} & & & \\
\midrule
\midrule
\multicolumn{4}{c}{\textbf{Base model: Llama3.2-VL-11B}} \\
\midrule
$T=2, \delta=0.1$ & 53.14 & 86.61 & 69.88 \\
$T=4, \delta=0.1$ & 52.93 & 86.37 & 69.65 \\
$T=8, \delta=0.1$ & 52.97 & 86.29 & 69.63 \\
$T=2, \delta=0.2$ & 53.08 & 85.87 & 69.48 \\
$T=2, \delta=0.3$ & 52.77 & 86.55 & 69.66 \\
\midrule
\midrule
\multicolumn{4}{c}{\textbf{Base model: Qwen-VL-Max}} \\
\midrule
$T=2, \delta=0.1$ & 76.40 & 97.41 & 86.91 \\
$T=4, \delta=0.1$ & 76.17 & 96.85 & 86.51 \\
$T=8, \delta=0.1$ & 75.93 & 96.79 & 86.36 \\
$T=2, \delta=0.2$ & 76.32 & 97.10 & 86.71 \\
$T=2, \delta=0.3$ & 76.18 & 97.33 & 86.76 \\
\bottomrule
\end{tabular}
\label{tab_hyperparameter}
\end{table}

\begin{table}
\centering
\caption{Results under various training settings using Llama3.2-VL-11B. ``RS'' denotes random seed, and ``\#40'' indicates that the number of training samples is 40.}
\setlength{\tabcolsep}{2.83pt}
\renewcommand{\arraystretch}{1.15}
\begin{tabular}{ccccc}
    \toprule
    \textbf{\makecell[c]{Data $\rightarrow$ \\ \textbf{Settings $\downarrow$}}} & \multirow{1}{*}{\textbf{MultimodalQA}} & \multirow{1}{*}{\textbf{Vidoseek}} & \multirow{1}{*}{\textbf{ScienceQA}} & \multirow{1}{*}{\textbf{Avg.}}   \\
    \midrule
    RS: 888 & 53.14 & 29.33 & 86.61 & 56.36 \\        
    RS: 40 & 52.58 & 28.73 & 86.59 & 55.97 \\       
    RS: 1234 & 52.64 & 28.55 & 85.82 & 55.67 \\  
    \midrule
    \#20 & 52.71 & 28.78 & 86.33 & 55.94 \\  
    \#40 & 53.14 & 29.33 & 86.61 & 56.36 \\
    \#80 & 52.85 & 28.66 & 85.90 & 55.80 \\
    \bottomrule
\end{tabular}
\label{training_settings}
\end{table}

\begin{table}
\small
\centering
\caption{Results with varying numbers of agents.} 
\setlength{\tabcolsep}{2.83pt}
\renewcommand{\arraystretch}{1.15}
\begin{tabular}{ccccc}
    \toprule
    \textbf{Dataset} & \textbf{MultimodalQA} & \textbf{Vidoseek} & \textbf{ScienceQA} & \textbf{Avg.} \\  
    \midrule
    Default setting & 53.14 & 29.33 & 86.61 & 56.36 \\        
    +1 agent &51.39 & 26.04 & 83.68 & 53.70 \\       
    -1 agent & 50.42 & 25.21 & 84.93 & 53.52 \\         
    \bottomrule
\end{tabular}
\label{diff_agent_number}
\end{table}

\begin{table*}[!t]
    \small
    \centering
    \caption{Performance comparison on the domain-specific ScienceQA benchmark using Qwen2.5-VL (7B).    }
    \setlength{\tabcolsep}{2.83pt}
    \renewcommand{\arraystretch}{0.85}
    \begin{tabular}{cc|ccc|ccc|cc|c}
        \toprule
        \multirow{3}*{\makecell{\textbf{Training} \\ \textbf{Paradigm}}}
        & \multirow{3}*{\textbf{Method}} & \multicolumn{3}{c}{\raisebox{-0.5em}{\textbf{Subject}}} & \multicolumn{3}{c}{\raisebox{-0.5em}{\textbf{Context Modality}}} & \multicolumn{2}{c|}{\raisebox{-0.5em}{\textbf{Grade}}} & \multirow{3}{*}{\makecell{\textbf{Avg.}}} \\
        & & \multicolumn{3}{c}{\rule{60pt}{0.4pt}} & \multicolumn{3}{c}{\rule{60pt}{0.4pt}} & \multicolumn{2}{c|}{\rule{40pt}{0.4pt}} \\ 
        & & \textbf{NAT} & \textbf{Soc} & \textbf{LAN} & \textbf{TXT} & \textbf{IMG} & \textbf{NO} & \textbf{G1-6} & \textbf{G7-12} \\
        \midrule

        \multirow{2}{*}{Zero-shot} & SP & 82.84 & 84.71 & 78.00 & 81.26 & 79.07 & 80.88 & 85.23 & 76.13 & 81.98$_{\pm 1.0}$ \\
        & CoT & 82.25 & 91.21 & 77.55 & 81.85 & 81.84 & 80.03 & 85.92 & 77.50 & 82.91$_{\pm 0.9}$ \\
        \cmidrule(lr){1-11}
        \multirow{5}{*}{\makecell{Single-agent \\ RAG}} & Wiki-LLaVA & 82.90 & 84.98 & 77.91 & 81.52 & 80.51 & 80.09 & 84.78 & 76.15 & 82.11$_{\pm 0.5}$ \\
        & RoRA-VLM & 83.37 & 86.65 & 78.84 & 82.61 & 80.97 & 81.53 & 86.41 & 77.59 & 82.88$_{\pm 1.3}$ \\
        & EchoSight & 82.97 & 86.04 & 78.27 & 82.18 & 80.67 & 80.72 & 85.62 & 76.61 & 82.40$_{\pm 0.7}$ \\
        & LLaVA-mR2AG & 83.45 & 86.90 & 79.23 & 82.95 & 81.04 & 81.96 & 86.71 & 77.83 & 82.99$_{\pm 0.5}$ \\
        & CoRe-MMRAG & 83.47 & 86.98 & 79.56 & 83.27 & 81.24 & \underline{82.06} & 86.93 & 77.90 & 83.10$_{\pm 1.5}$ \\
        \cmidrule(lr){1-11}
        \multirow{5}{*}{\makecell{Multi-agent \\ RAG}} & OmniSearch & 81.78 & \textbf{92.38} & \underline{80.00} & 81.29 & 81.42 & 81.63 & 86.11 & 78.92 & 83.54$_{\pm 0.5}$ \\
        & ViDoRAG & \underline{86.77} & 84.28 & 76.82 & \underline{85.68} & 84.12 & 79.60 & 86.33 & 78.88 & 83.66$_{\pm 0.9}$ \\
        & HM-RAG & 83.65 & \underline{92.37} & \textbf{80.27} & 83.57 & \underline{85.67} & 80.83 & 85.94 & \textbf{82.19} & \underline{84.60}$_{\pm 1.1}$ \\
        & E-Agent & 85.30 & 87.40 & 78.27 & 84.95 & 83.14 & 80.84 & \underline{87.04} & 78.31 & 83.92$_{\pm 1.0}$ \\
       \rowcolor[HTML]{C0C0C0} & \textbf{Ours} & \textbf{90.63} & 89.65 & \underline{80.00} & \textbf{89.00} & \textbf{88.80} & \textbf{82.79} & \textbf{90.90} & \underline{81.87} & \textbf{87.67}$_{\pm 0.7}$ \\
        \bottomrule
    \end{tabular}
    \label{main_exp_scienceQA}
\end{table*}

\begin{table*}[!tb]
    \scriptsize
    \centering
    \caption{Performance comparison on general-domain benchmarks using Qwen2.5-VL (7B). }
    \setlength{\tabcolsep}{1.8pt}
    \renewcommand{\arraystretch}{0.85}
    \begin{tabular}{cc|cccccccccccc|cccc|cc}
        \toprule
        \multirow{5}*[-0.85em]{\makecell{\textbf{Training} \\ \textbf{Paradigm}}}
        & \multirow{5}*[-0.85em]{\textbf{Method}} & \multicolumn{12}{c|}{\raisebox{-0.5em}{\textbf{Vidoseek}}} & \multicolumn{4}{c|}{\raisebox{-0.5em}{\textbf{MultimodalQA}}} & \multicolumn{2}{c}{\multirow{4}{*}{\textbf{Average}}} \\
        & & \multicolumn{12}{c|}{\rule{226pt}{0.4pt}}  & \multicolumn{4}{c|}{\rule{70pt}{0.4pt}} \\
        & & \multicolumn{2}{c|}{\raisebox{-0.5em}{\textbf{Single-hop}}} & \multicolumn{2}{c|}{\raisebox{-0.5em}{\textbf{Multi-hop}}} & \multicolumn{2}{c|}{\raisebox{-0.5em}{\textbf{Text}}} & \multicolumn{2}{c|}{\raisebox{-0.5em}{\textbf{Table}}} & \multicolumn{2}{c|}{\raisebox{-0.5em}{\textbf{Chart}}} & \multicolumn{2}{c|}{\raisebox{-0.5em}{\textbf{Layout}}} &  \multicolumn{2}{c|}{\raisebox{-0.5em}{\textbf{Image}}} & \multicolumn{2}{c|}{\raisebox{-0.5em}{\textbf{Text}}} \\
       & & \multicolumn{2}{c}{\rule{33pt}{0.4pt}} & \multicolumn{2}{c}{\rule{33pt}{0.4pt}} & \multicolumn{2}{c}{\rule{33pt}{0.4pt}} & \multicolumn{2}{c}{\rule{33pt}{0.4pt}} & \multicolumn{2}{c}{\rule{33pt}{0.4pt}} & \multicolumn{2}{c|}{\rule{33pt}{0.4pt}} & \multicolumn{2}{c}{\rule{33pt}{0.4pt}} & \multicolumn{2}{c|}{\rule{33pt}{0.4pt}} & \multicolumn{2}{c}{\rule{71pt}{0.4pt}} \\
       & & \textbf{Acc$^\star$} & \textbf{EM} & \textbf{Acc$^\star$} & \textbf{EM} & \textbf{Acc$^\star$} & \textbf{EM} & \textbf{Acc$^\star$} & \textbf{EM} & \textbf{Acc$^\star$} & \textbf{EM} & \textbf{Acc$^\star$} & \textbf{EM} & \textbf{Acc$^\star$} & \textbf{EM} & \textbf{Acc$^\star$} & \textbf{EM} & \textbf{Acc$^\star$} & \textbf{EM} \\
       \midrule

        \multirow{2}{*}{Zero-shot} & SP & 26.51 & 2.79 & 11.47 & 6.44 & 25.00 & 1.25 & 10.86 & 4.57 & 15.29 & 12.10 & 22.60 & 3.01 & 25.50 & 23.23 & 32.94 & 26.48 & 21.27$_{\pm 1.1}$ & 9.98$_{\pm 1.0}$ \\
        & CoT & 26.58 & 3.86 & 11.66 & 8.84 & 28.50 & 2.00 & 13.57 & 7.14 & 15.65 & 14.10 & 21.68 & 4.47 & 23.86 & 20.23 & 36.20 & 29.24 & 22.21$_{\pm 0.8}$ & 11.24$_{\pm 0.9}$ \\
        \midrule
        \multirow{5}{*}{\makecell{Single-agent \\ RAG}} & Wiki-LLaVA & 51.56 & 17.34 & 47.87 & 39.12 & 53.99 & 9.24 & 41.68 & 32.75 & 43.24 & 36.75 & 53.12 & 25.37 & 20.37 & 18.11 & 62.67 & 53.32 & 46.81$_{\pm 0.8}$ & 29.00$_{\pm 1.2}$ \\
        & RoRA-VLM & 52.20 & 17.83 & 47.92 & 39.18 & 54.86 & 9.69 & 42.18 & 33.02 & 43.89 & 36.90 & 53.21 & 25.51 & 20.68 & 18.12 & 62.88 & 53.39 & 47.23$_{\pm 0.6}$ & 29.21$_{\pm 0.9}$ \\
        & EchoSight & 52.25 & 17.98 & 48.09 & 39.24 & 55.00 & 10.00 & 42.29 & 33.14 & 43.95 & 36.94 & 53.29 & 25.62 & 21.00 & 18.18 & 62.96 & 53.48 & 47.35$_{\pm 0.5}$ & 29.32$_{\pm 0.4}$ \\
        & LLaVA-mR2AG & 52.76 & 18.24 & 48.47 & 39.67 & 55.54 & 10.34 & 42.67 & 33.56 & 44.23 & 37.25 & 53.32 & 25.94 & 21.23 & 18.54 & 63.12 & 53.66 & 47.67$_{\pm 0.8}$ & 29.65$_{\pm 0.7}$ \\
        & CoRe-MMRAG & 52.51 & 18.17 & 48.26 & 39.59 & 55.25 & 10.18 & 42.49 & 33.32 & 44.06 & 37.18 & 53.28 & 25.81 & 21.11 & 18.39 & 63.03 & 53.50 & 47.50$_{\pm 0.5}$ & 29.52$_{\pm 0.9}$ \\
        \midrule
        \multirow{5}{*}{\makecell{Multi-agent \\ RAG}} & OmniSearch & 53.10 & \underline{20.26} & 49.44 & 37.99 & 53.75 & 17.25 & 44.86 & 31.71 & 48.22 & 38.48 & 53.56 & 26.00 & 28.86 & 26.36 & 61.01 & 55.82 & 49.10$_{\pm 1.2}$ & 31.73$_{\pm 0.8}$ \\
        & ViDoRAG &56.16 & 17.80 & \underline{60.73} & \underline{49.47} & \underline{73.75} & \underline{19.00} & \underline{57.00} & \underline{45.14} & \underline{57.23} & \underline{49.22} & 56.92 & 25.92 & 25.95 & 21.73 & \underline{72.18} & \underline{62.23} & \underline{57.49}$_{\pm 0.5}$ & \underline{36.31}$_{\pm 0.9}$ \\
        & HM-RAG & 59.42 & 19.02 & 51.28 & 41.61 & 60.00 & 17.25 & 47.29 & 36.86 & 44.49 & 39.12 & \underline{59.93} & 26.00 & 30.86 & 26.82 & 68.84 & 58.23 & 52.76$_{\pm 0.7}$ & 33.11$_{\pm 1.2}$ \\
        & E-Agent & \underline{59.49} & 19.19 & 53.69 & 42.63 & 66.00 & 16.50 & 50.00 & 37.71 & 48.04 & 40.31 & 59.56 & \underline{26.47} & \underline{37.82} & \underline{33.23} & 69.29 & 60.09 & 55.49$_{\pm 1.1}$ & 34.52$_{\pm 0.8}$ \\
         \rowcolor[HTML]{C0C0C0} & \textbf{Ours} & \textbf{63.57} & \textbf{24.19} & \textbf{66.20} & \textbf{55.13} & \textbf{75.00} & \textbf{26.25} & \textbf{62.86} & \textbf{50.86} & \textbf{57.96} & \textbf{52.87} & \textbf{65.48} & \textbf{32.47} & \textbf{59.32} & \textbf{58.18} & \textbf{77.97} & \textbf{70.51} & \textbf{66.05}$_{\pm 0.7}$ & \textbf{46.31}$_{\pm 0.6}$ \\

        \bottomrule
    \end{tabular}
    
    \label{main_exp_general_datasets}
\end{table*}

\section{Other Experiments}
\label{other_exp}

\begin{figure}[!t]
\centering
\includegraphics[width=7.5cm]{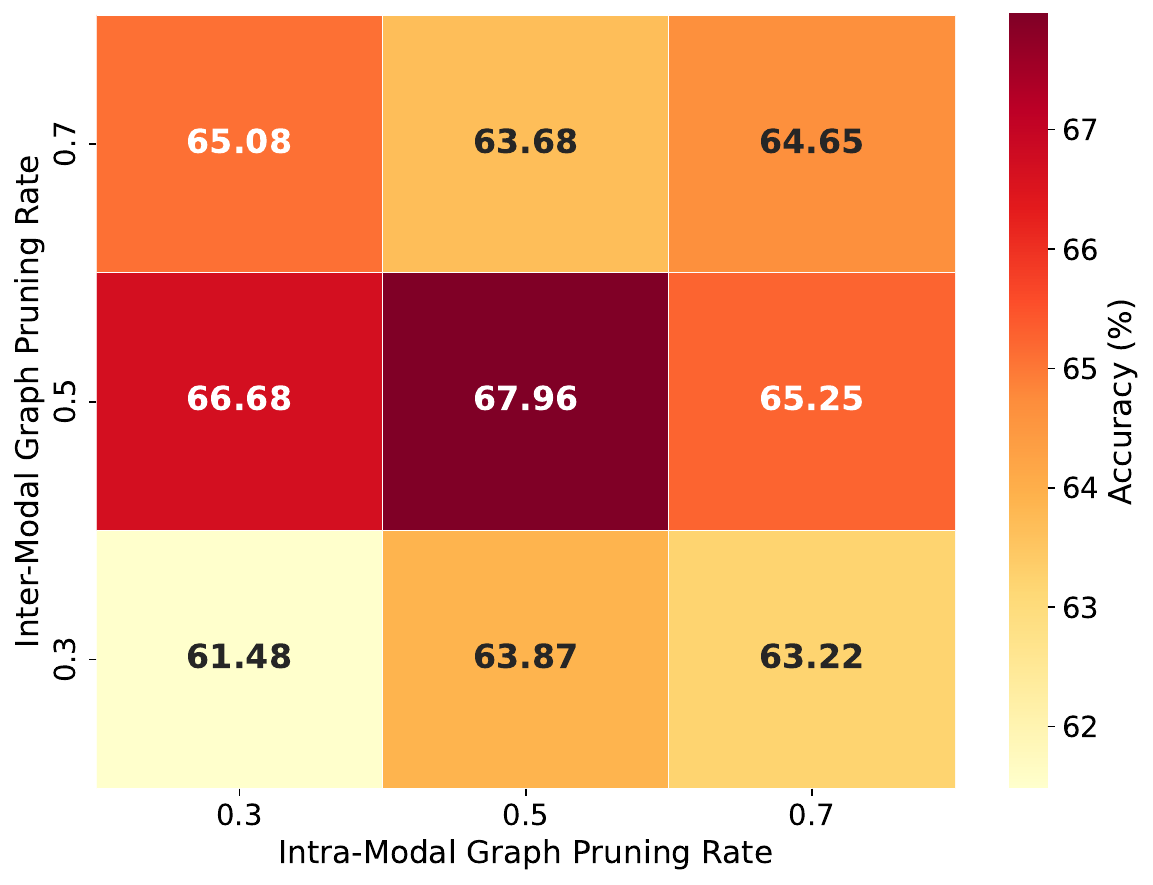}
\caption{The influence of different edge pruning rates using Qwen2.5-VL-7B.
}
\label{pruning_rate_qwen}
\end{figure}

\begin{figure}[!t]
\centering
\includegraphics[width=7.5cm]{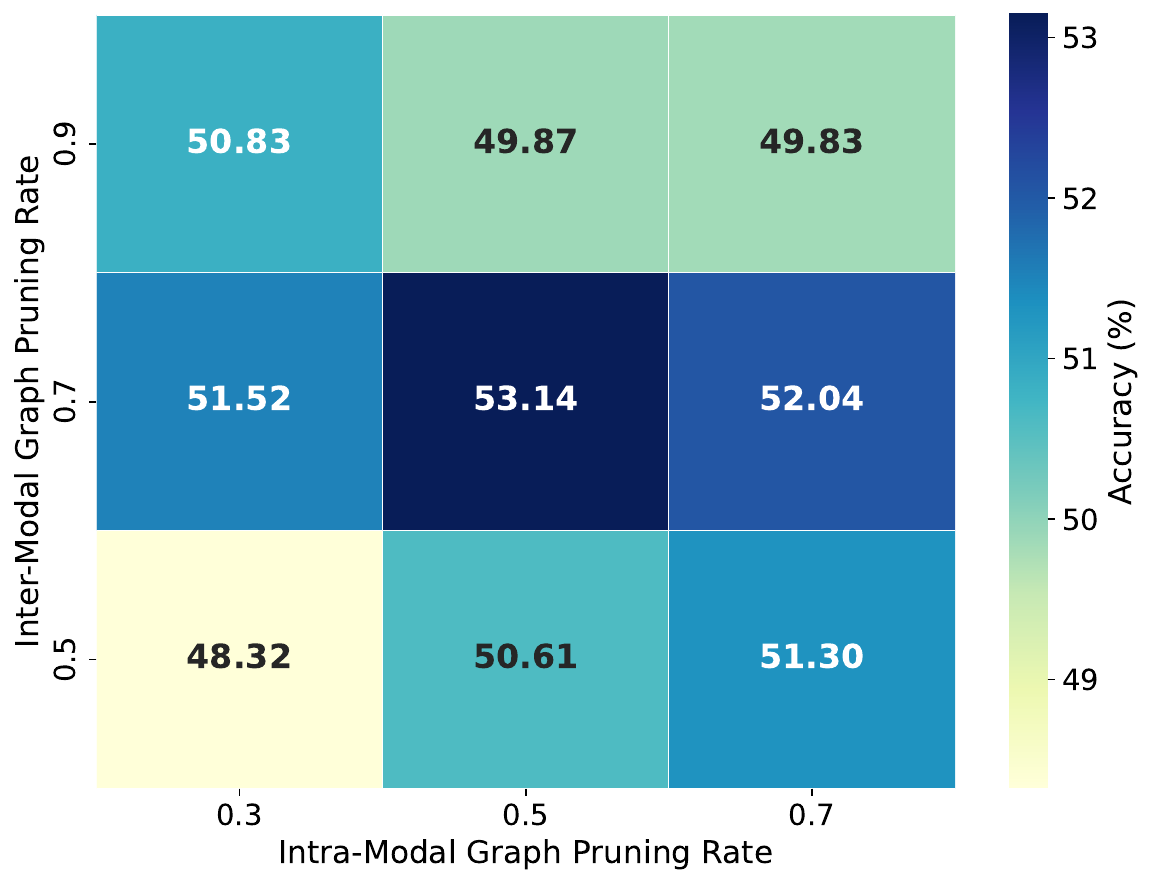}
\caption{The influence of different edge pruning rates using Llama3.2-VL-11B.
}
\label{pruning_rate_llama}
\end{figure}

\subsection{General Performance}
\label{general_perf}
Table~\ref{main_exp_scienceQA} and Table~\ref{main_exp_general_datasets} present the performance of Qwen2.5-VL-7B on ScienceQA, Vidoseek, and MultimodalQA.
We observe that our model achieves state-of-the-art performance on all three datasets, with the most substantial improvement on MultimodalQA.

\subsection{Hyperparameter Analysis}
\label{hyper_para}
\begin{itemize}
    \item \textbf{Hyperparameters:} Table~\ref{tab_hyperparameter} reports performance under varying numbers of communication rounds ($T$) and noise levels ($\delta$).
    Experiments on Llama3.2-VL-11B and Qwen-VL-Max show that performance remains stable across these settings.
    Table~\ref{training_settings} presents experiments with different numbers of training samples and random seeds to demonstrate stability.
    
    \item \textbf{Varying Number of Agents:} We adapt the number of agents to the multi-modal complexity and task requirements of each dataset.
    For example, MultimodalQA, which emphasizes textual reasoning with relatively lightweight visual cues, uses 5 textual and 4 visual agents.
    We further investigate the impact of agent configuration by adding or removing one textual agent and one visual agent for each dataset.
    The choice of the number of agents is based on dataset complexity and task requirements, such as MultimodalQA relying more heavily on textual information, and is determined empirically.
    Table~\ref{diff_agent_number} shows that our model is insensitive to small changes in the number of agents and exhibits a certain degree of robustness.
    
    \item \textbf{Pruning Rates Analysis:} To study how pruning strength affects M$^3$Prune, we evaluate intra- and inter-modal pruning rates under varying sparsity configurations on MultimodalQA, using Qwen2.5-VL-7B and Llama3.2-VL-11B as backbone models.
    We sweep the pruning rates from very low to very high to assess behavior under overly dense and overly sparse topologies.
    As illustrated in Fig.~\ref{pruning_rate_qwen} and Fig.~\ref{pruning_rate_llama}, both overly conservative and overly aggressive pruning yield suboptimal performance.
    Excessively low pruning rates retain redundant edges, introducing noise that degrades decision quality, whereas excessively high pruning rates remove critical communication pathways.
    Optimal performance arises at moderate pruning levels, indicating that balanced sparsity preserves informative interactions while filtering noise.
\end{itemize}


\begin{figure}[!t]
\centering
\includegraphics[width=7.5cm]{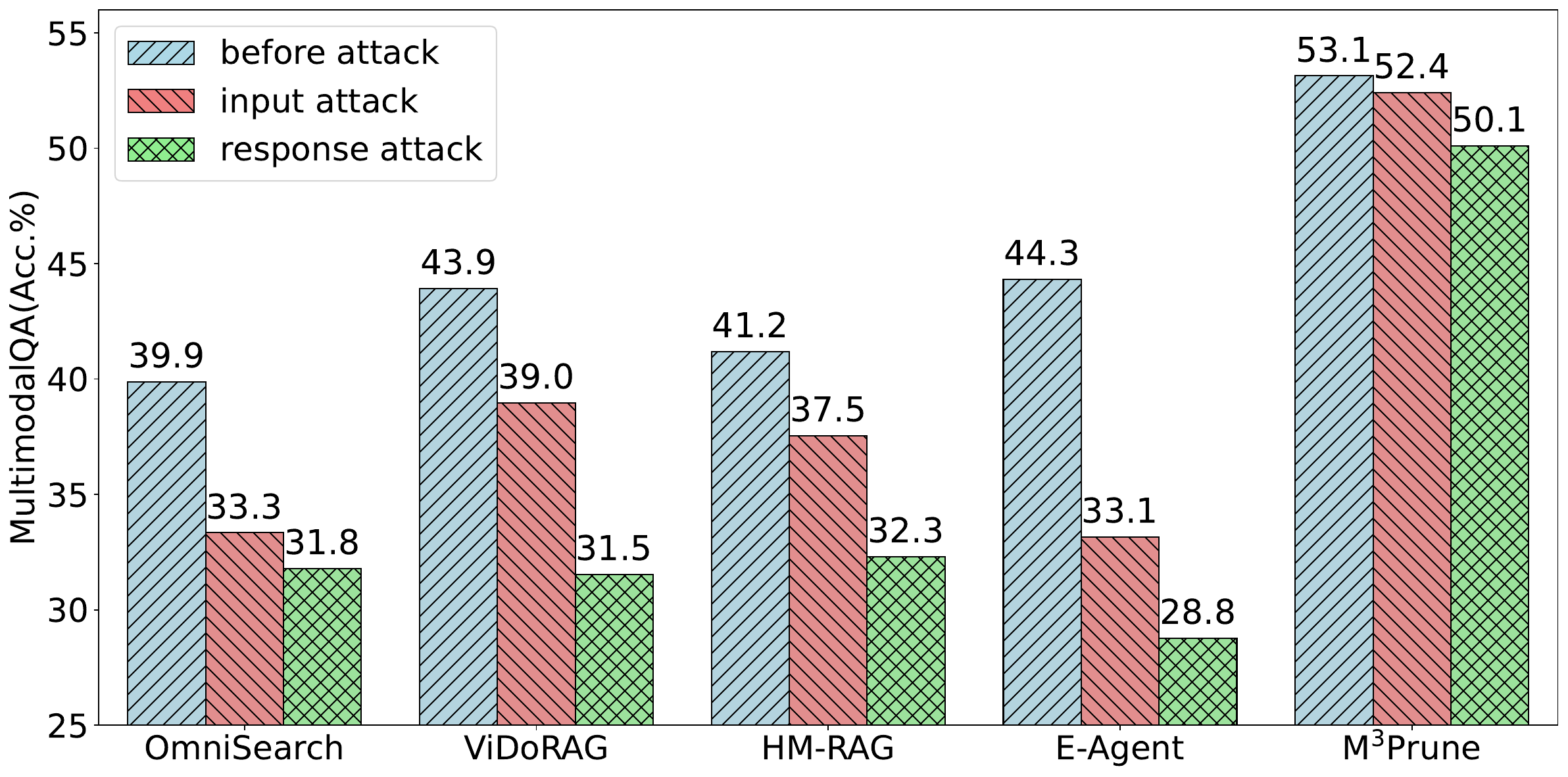}
\caption{Performance under adversarial attacks, including input prompt and response perturbations on MultimodalQA.}
\label{robustness_attack_llama}
\end{figure}

\begin{figure*}[!t]
\centering
\includegraphics[width=17.5cm, height=9.5cm]{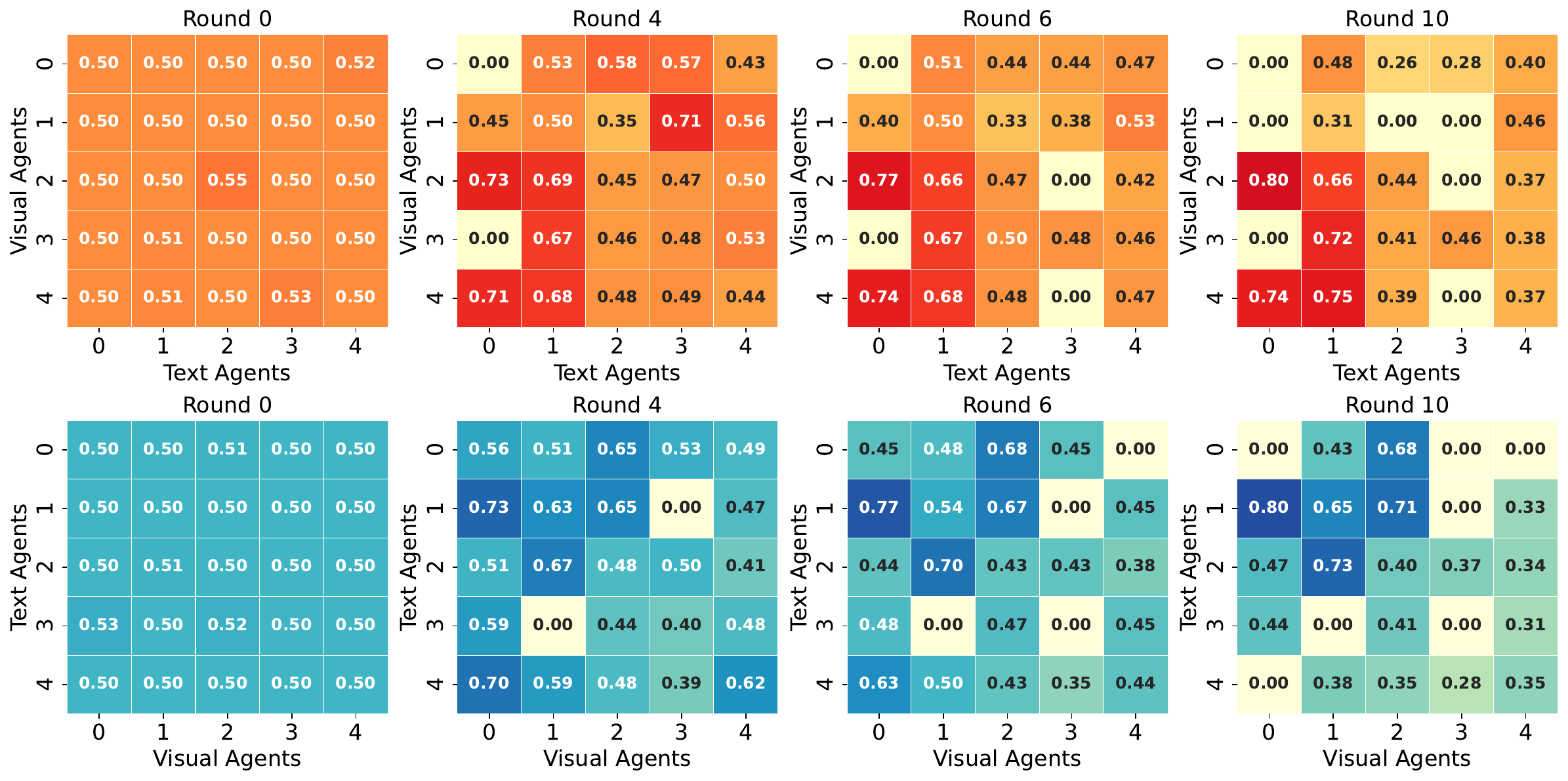}
\caption{Evolution of communication edge weights from visual-to-text (Top) and text-to-visual (Bottom) directions on ScienceQA.}
\label{complete_edge_weight_scienceqa}
\end{figure*}

\begin{figure*}[!t]
\centering
\includegraphics[width=17.5cm, height=9.5cm]{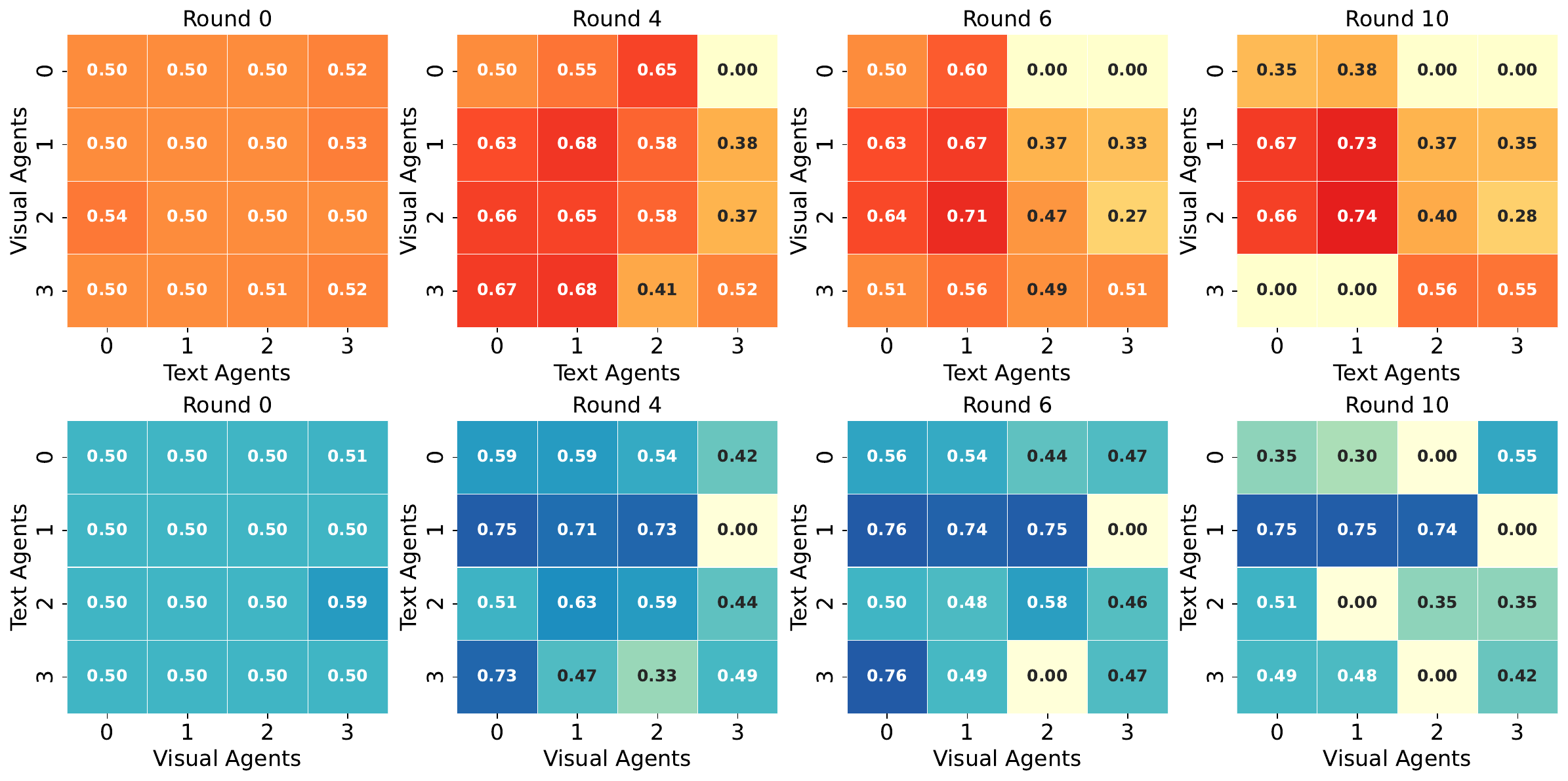}
\caption{Evolution of communication edge weights from visual-to-text (Top) and text-to-visual (Bottom) directions on Vidoseek.}
\label{complete_edge_weight_vidoseek}
\end{figure*}

\subsection{Communication Edge Evolution}
\label{sec_complete_edge}
Fig.~\ref{complete_edge_weight_scienceqa} and Fig.~\ref{complete_edge_weight_vidoseek} illustrate the full evolution of edge weights on ScienceQA and Vidoseek using Qwen-VL-Max.
Initially, the softened adjacency scores between all agents are approximately 0.5.
As the number of dialogue rounds increases, high edge weights become concentrated on a few key agent pairs, while the weights between other agents gradually decrease to the range of 0.2--0.5.

\begin{figure*}[!t]
\centering
\includegraphics[width=17cm, height=5.5cm]{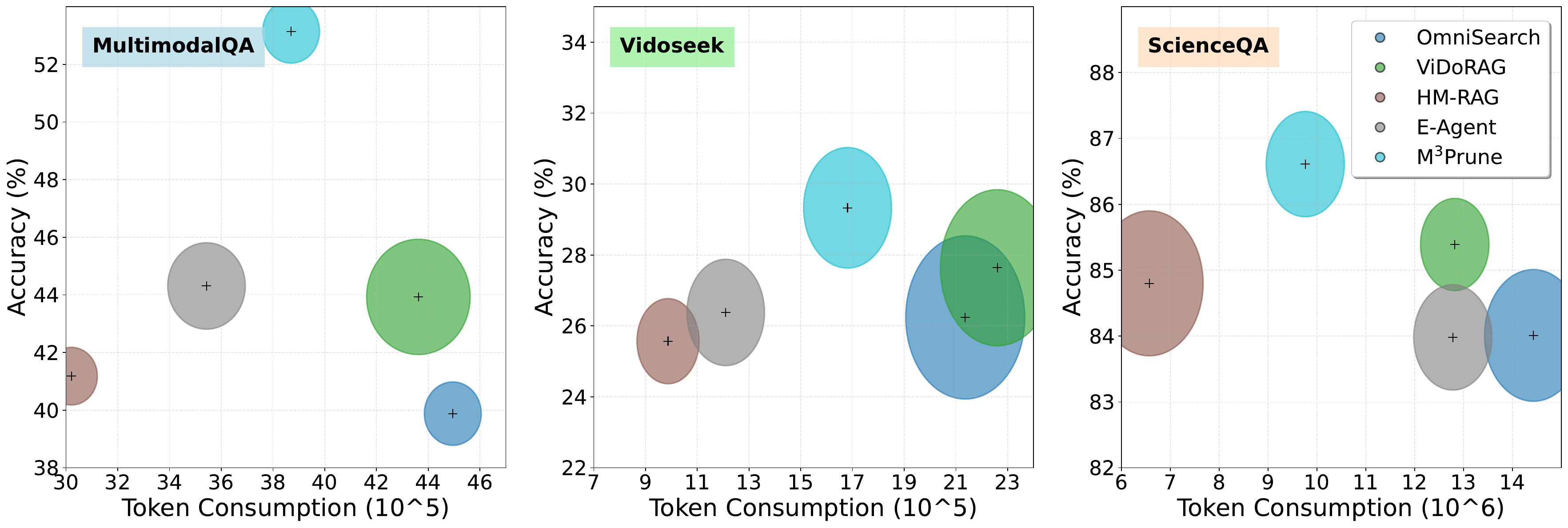}
\caption{Token efficiency comparison for different multi-agent models on Llama3.2-VL-11B. The number of tokens consumed is calculated by the sum of prompt tokens and completion tokens.}
\label{token_performance_llama}
\end{figure*}

\begin{figure*}[!t]
\centering
\includegraphics[width=17cm, height=5.5cm]{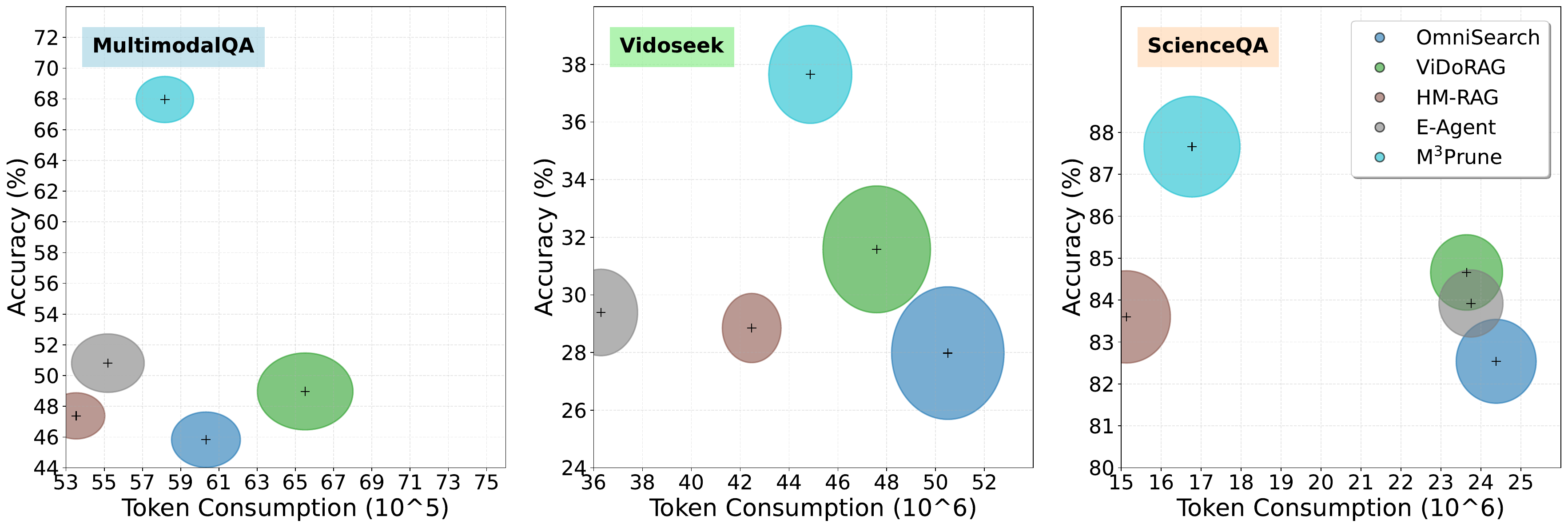}
\caption{Token efficiency comparison for different multi-agent models on Qwen2.5-VL-7B.}
\label{token_performance_qwen}
\end{figure*}

\subsection{Token Efficiency}
\label{sec_token_comp}
The overall token efficiency is illustrated in Fig.~\ref{token_performance_llama} and Fig.~\ref{token_performance_qwen}.
M$^3$Prune achieves superior performance across various backbone models and datasets while maintaining moderate token usage.
Notably, it yields the largest performance improvements on the MultimodalQA dataset and the lowest token consumption on ScienceQA.

\subsection{Robustness Verification}
The robustness verification experiment on MultimodalQA using Llama3.2-VL-11B is presented in Fig.~\ref{robustness_attack_llama}.
Consistent with the results on Qwen2.5-VL-7B, M$^3$Prune exhibits substantially smaller performance degradation under both attack types than the other models.

\subsection{Case Study}
\label{case_study}

We present a case study on the ScienceQA dataset to illustrate the intra- and inter-modal collaboration among agents.
Initially, the Text Knowledge Expert identifies the key entities, while the Image Data Analyst integrates the contextual information with the image and notes that the first step is to recognize the seven continents.
Building on this observation, the Text Critic emphasizes the importance of identifying the highlighted areas in the image.
The Image Scientist then confirms that recognizing these highlighted regions is crucial for answering the question.
Next, the Text Scientist synthesizes the inputs from the previous agents and concludes that the dark green areas correspond to the highlighted regions.
In the second dialogue round, the Text and Image Data Analysts leverage their respective expertise to identify the highlighted area as Australia.
Based on this joint reasoning, the Text and Image Scientists determine scientifically that the correct answer is \textbf{B}.
Finally, the Image and Text Critics evaluate and validate the analyses provided by all preceding agents.

\begin{figure*}[!t]
\centering
\includegraphics[width=16cm]{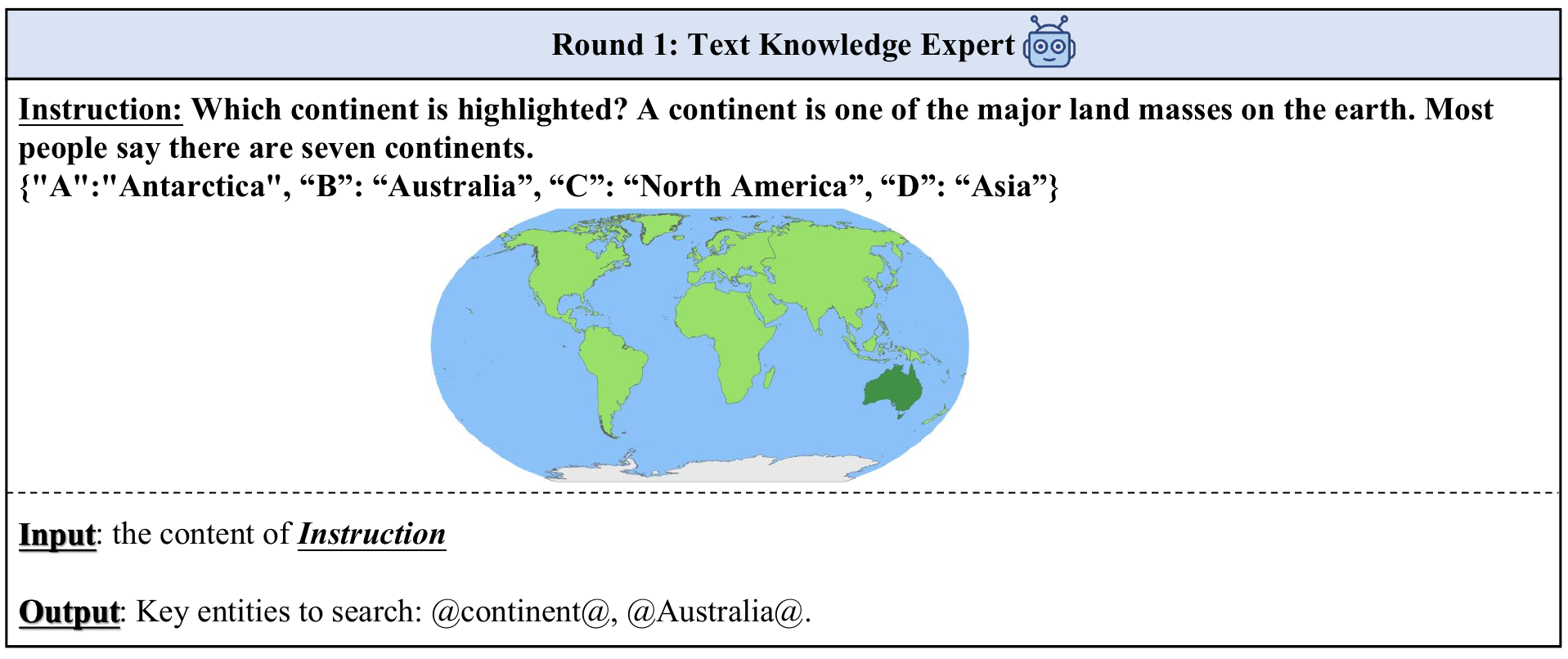}
\vspace{-0.25cm}
\end{figure*}

\begin{figure*}[!t]
\centering
\includegraphics[width=16cm]{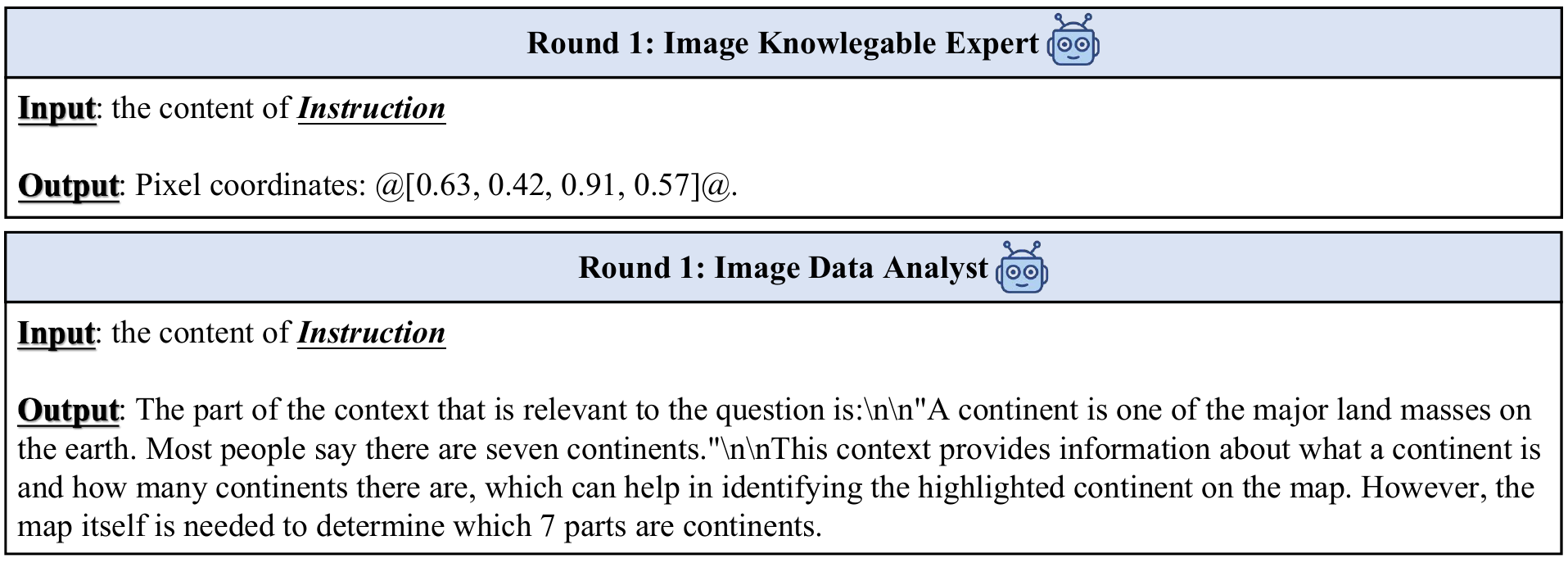}
\vspace{-0.25cm}
\end{figure*}

\begin{figure*}[!t]
\centering
\includegraphics[width=16cm]{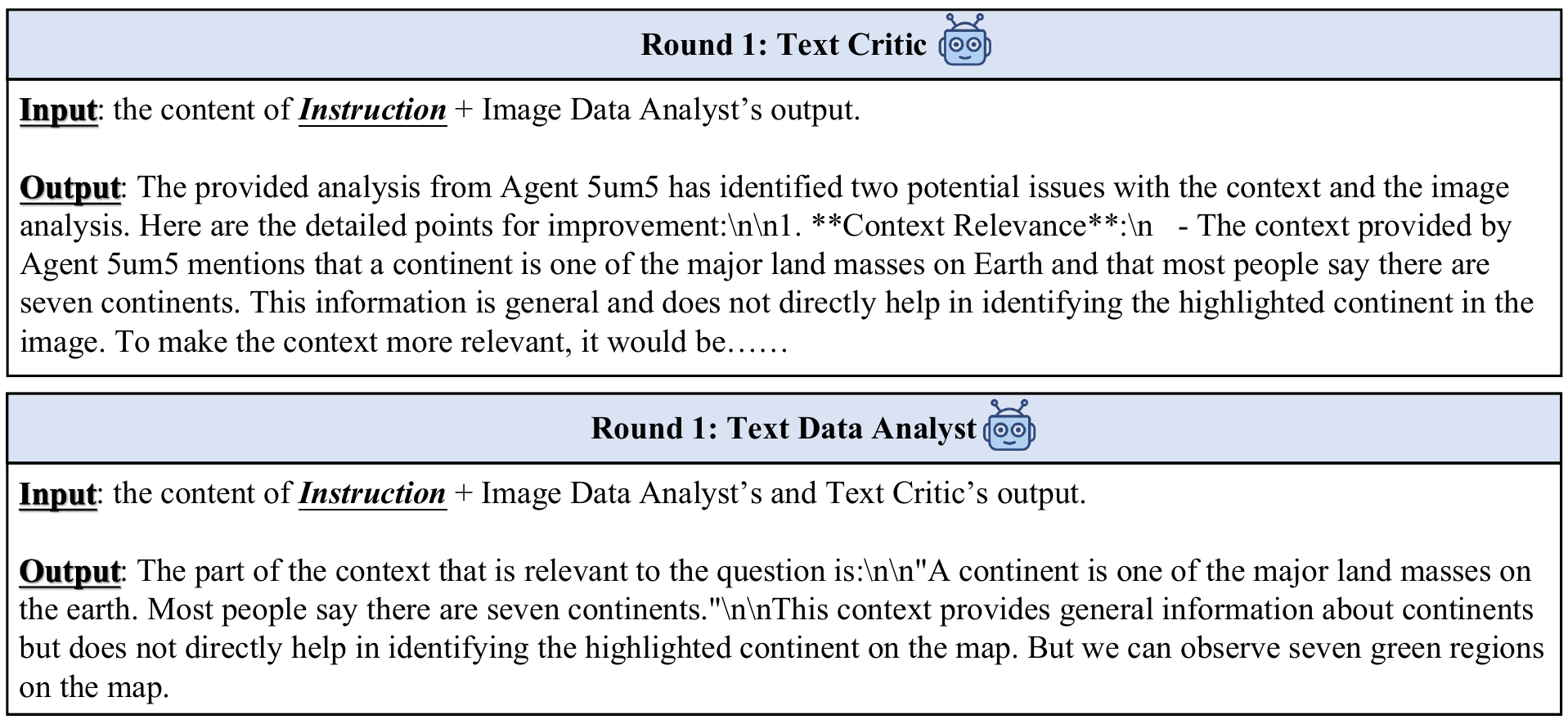}
\vspace{-0.25cm}
\end{figure*}

\begin{figure*}[!t]
\centering
\includegraphics[width=16cm]{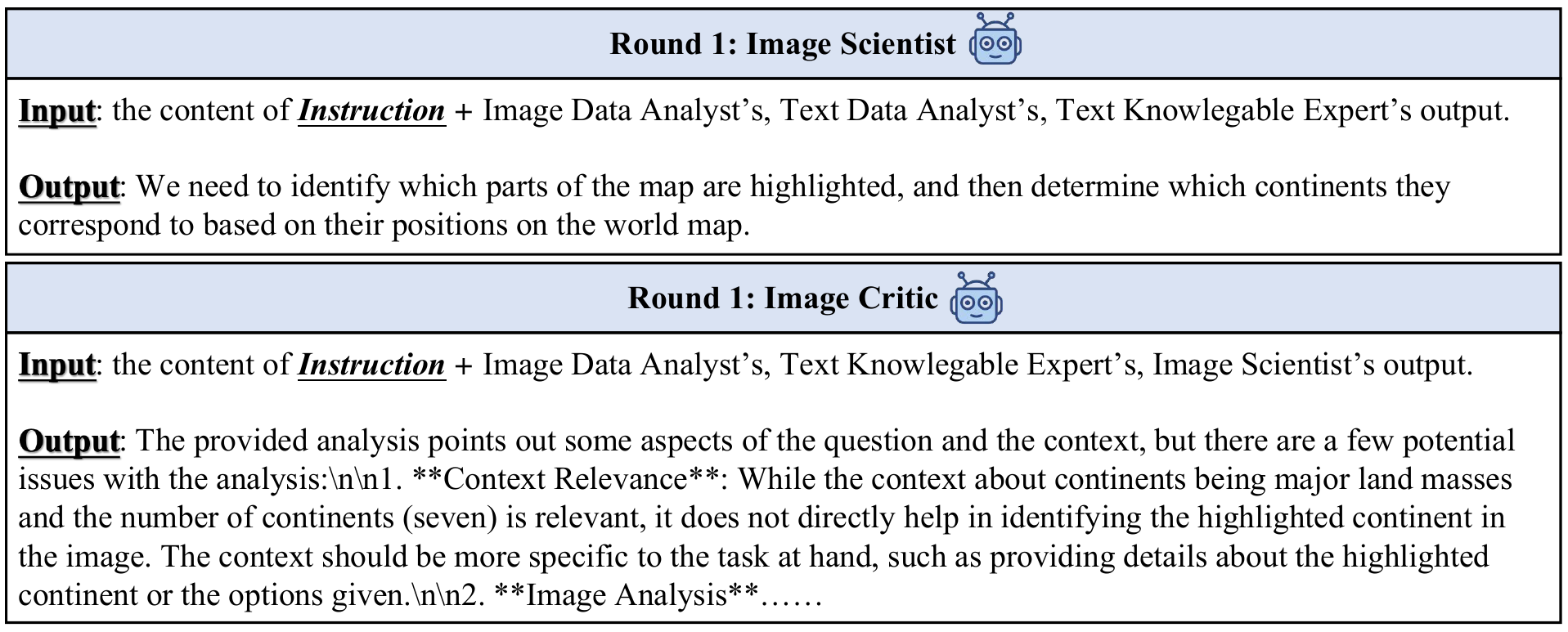}
\vspace{-0.25cm}
\end{figure*}

\begin{figure*}[!t]
\centering
\includegraphics[width=16cm]{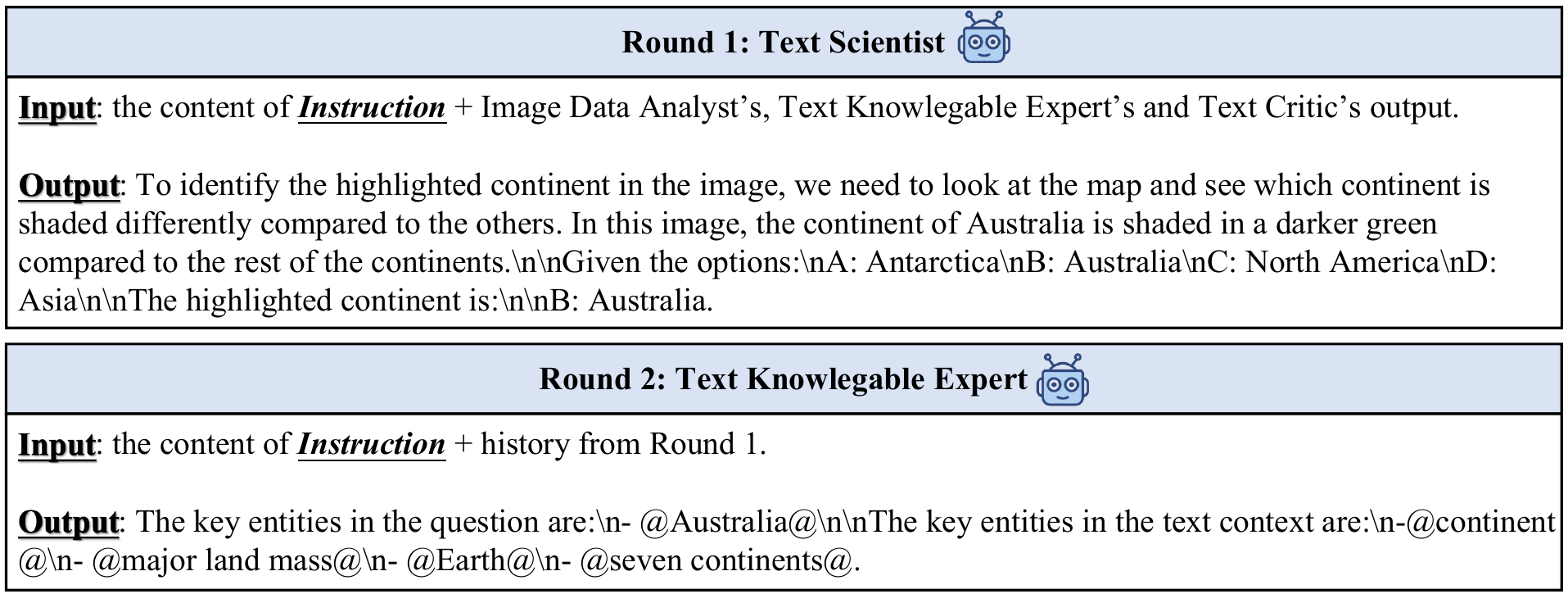}
\vspace{-0.25cm}
\end{figure*}

\begin{figure*}[!t]
\centering
\includegraphics[width=16cm]{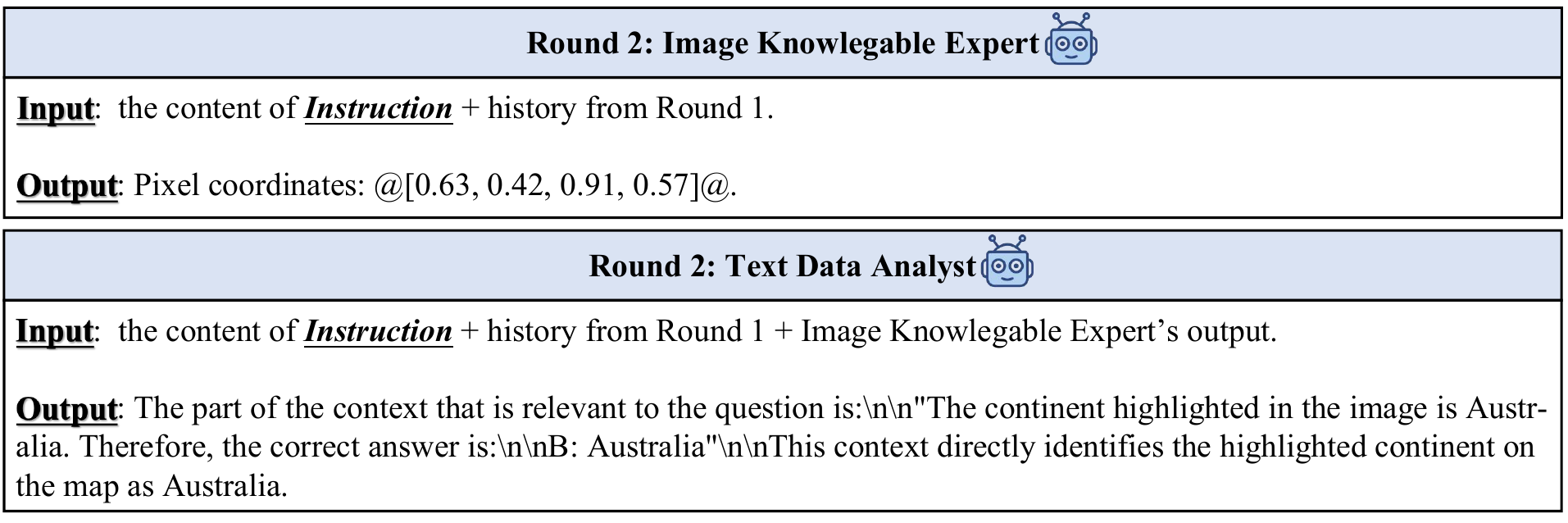}
\vspace{-0.25cm}
\end{figure*}

\begin{figure*}[!t]
\centering
\includegraphics[width=16cm]{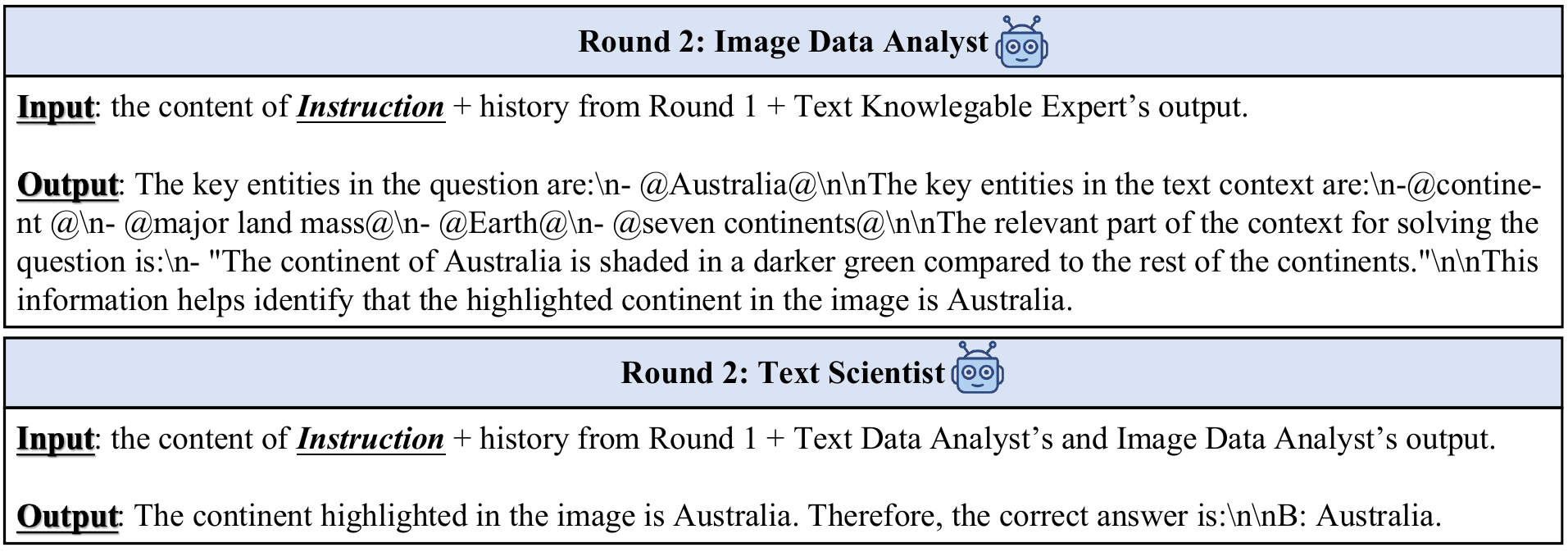}
\vspace{-0.25cm}
\end{figure*}

\begin{figure*}[!t]
\centering
\includegraphics[width=16cm]{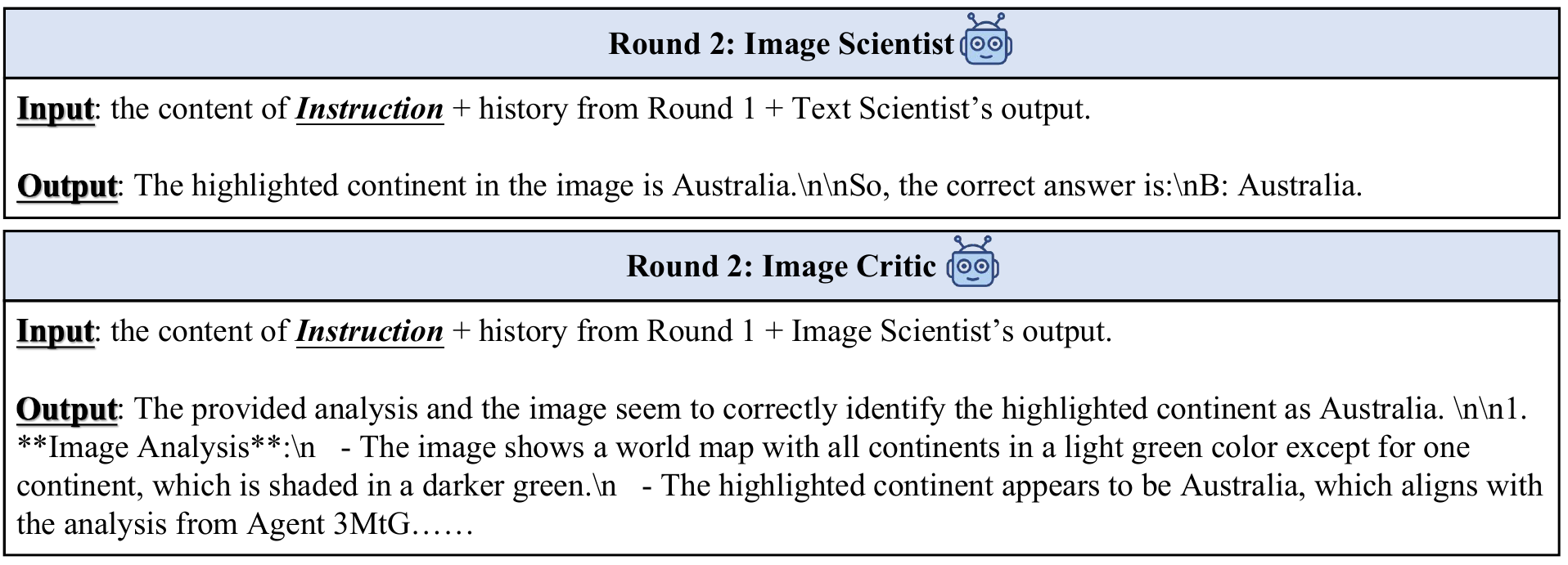}
\vspace{-0.25cm}
\end{figure*}

\begin{figure*}[!t]
\centering
\includegraphics[width=16cm]{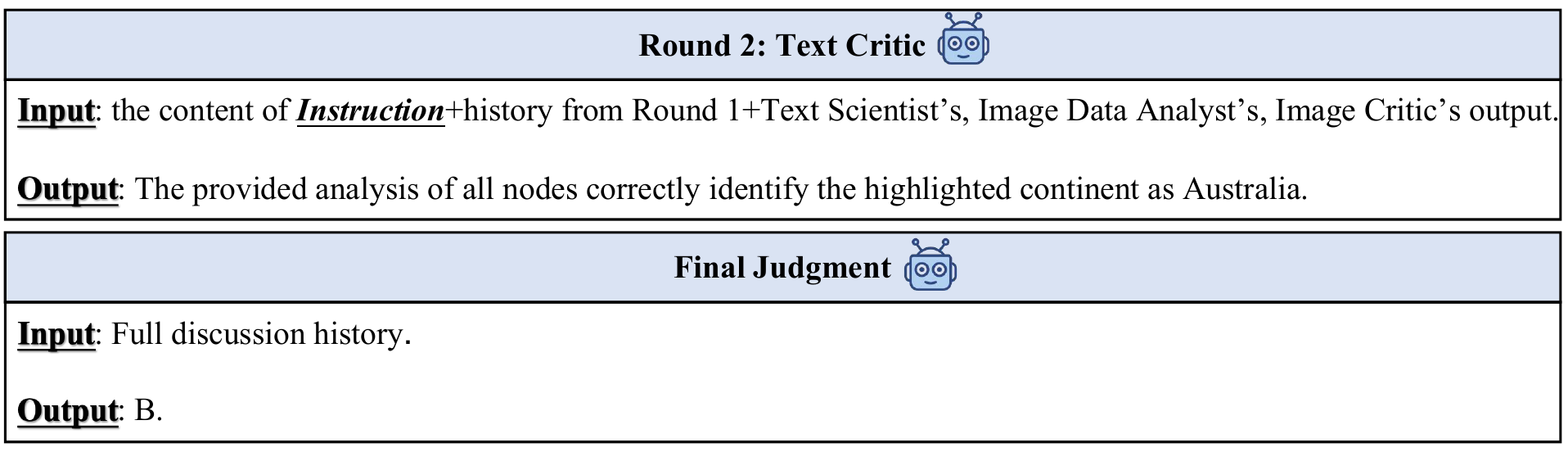}
\vspace{-0.25cm}
\end{figure*}
\end{document}